\documentclass{aa}

\usepackage{graphicx}
\usepackage{txfonts}
\usepackage{lscape}             
\usepackage{placeins}           
\usepackage{orcidlink}          
\usepackage{hyperref}

\newcommand{\system}{V1216\,Sco}

\newcommand{\Msun}{M$_{\odot}$} 
\newcommand{\cpd}{d$^{-1}$} 

\newcommand{\WD}{\textsc{WD}}
\newcommand{\TESS}{\textit{TESS }}
\newcommand{\tylde}{\raisebox{0.5ex}{\texttildelow}}

\newcommand\T{\rule{0pt}{2.6ex}}       
\newcommand\B{\rule[-1.2ex]{0pt}{0pt}} 

\usepackage{color}
\definecolor{todo}{rgb}{0.89,0.0,0.13}

\usepackage{color}
\definecolor{review}{RGB}{191, 66, 245}

\usepackage{color}
\definecolor{green(new)}{RGB}{50, 200, 60}

\usepackage{color}
\definecolor{cadetgrey}{rgb}{0.57, 0.64, 0.69}

\usepackage{color}
\definecolor{oror}{RGB}{0,150,0}

\newcommand{\comment}[1]{\iffalse{#1}\fi}

\usepackage{amsmath}

\begin{document}

\title{Eclipsing binary systems with $\beta$\,Cephei components}
\subtitle{V1216\,Sco}

\author{A. Miszuda \orcidlink{0000-0002-9382-2542} \inst{\ref{CAMK}} \and
        C. I. Eze \orcidlink{0000-0003-3119-0399} \inst{\ref{CAMK},\ref{Nigeria}} \and
        F. Kahraman Ali\c cavu\c s \orcidlink{0000-0002-9036-7476} \inst{\ref{Turkey1},\ref{Turkey2}} \and
        C. Johnston \orcidlink{0000-0002-3054-4135} \inst{\ref{MPI},\ref{UK},\ref{Leuven}} \and
        G. Handler \orcidlink{0000-0001-7756-1568} \inst{\ref{CAMK}}}

\institute{Nicolaus Copernicus Astronomical Centre, Polish Academy of Sciences, Bartycka 18, PL-00-716 Warsaw, Poland \\
           \email{amiszuda@camk.edu.pl, miszuda.amadeusz.astronomy@gmail.com}\label{CAMK} \and
           Department of Physics and Astronomy, Faculty of Physical Sciences, University of Nigeria, Nsukka, Nigeria\label{Nigeria} \and
           \c Canakkale Onsekiz Mart University, Faculty of Sciences, Physics Department, TR-17100 \c Canakkale, T\"urkiye\label{Turkey1} \and
           \c Canakkale Onsekiz Mart University, Astrophysics Research Center and Ulupınar Observatory, TR-17100 \c Canakkale, T\"urkiye\label{Turkey2} \and
           Max-Planck-Institut f\"ur Astrophysik, Karl-Schwarzschild-Stra{\ss}e 1, G-85741 Garching bei M\"unchen, Germany\label{MPI} \and
           Astrophysics group, Department of Physics, University of Surrey, Guildford, GU2 7XH, United Kingdom\label{UK} \and
           Institute of Astronomy, KU Leuven, Celestijnenlaan 200D, B-3001 Leuven, Belgium\label{Leuven}
           }

\abstract{
We present a detailed analysis of the high-mass binary system V1216\,Sco, an eclipsing Algol-type binary hosting a $\beta$\,Cep pulsator, with an orbital period of 3.92\,days. This system was analysed using \TESS photometry and high-resolution spectroscopy from SALT HRS to investigate its orbital parameters, stellar properties, and evolutionary history. The \TESS light curve, comprising over 12\,000 data points, revealed 5 independent pulsation frequencies within the $\beta$\,Cep range of 5 to 7\,\cpd. 

Spectroscopic analysis provided radial velocities and disentangled atmospheric parameters, enabling precise orbital and evolutionary modelling. The system features a primary star of 11.72\,\Msun\ and a secondary of 4.34\,\Msun, the latter with a radius near its Roche lobe, indicating recent or ongoing mass transfer. 
Evolutionary modelling with \textsc{mesa-binary} suggests that V1216\,Sco went through case A mass transfer scenario, where mass transfer began while the donor was still on the main sequence. The system’s evolutionary models indicate an age of 15–30\,Myr, highlighting the significant impact of binary interactions on stellar evolution. 

This study underscores the value of combined observational and theoretical approaches to understanding complex systems like V1216\,Sco, emphasizing the role of mass transfer in shaping binary star evolution.
}


\maketitle

\section{Introduction} 
\label{sec:introduction}


Binary systems, particularly those showing eclipses, serve as robust tools for testing stellar evolution theory by providing high precision estimates of fundamental stellar parameters \citep{Prsa2016,Maxted2020}. When compared to grids of stellar evolution models, the precise stellar parameters derived from the analysis of eclipsing binaries are critical for determining the age of such systems, tracing their evolutionary history, and defining the initial conditions that led them to their current state \citep[e.g.,][]{Rosales2019,Miszuda2021,Miszuda2022,Celik2024}. The accuracy of these measurements is further enriched when binary systems also contain pulsating components. This emerging branch of astrophysics simultaneously utilises information from binary modelling and asteroseismology to both cross calibrate evolution models \citep{Johnston2019,Beck2018,Sekaran2021} as well as place constraints on the impact of binary evolution on asteroseismic signals \citep{Murphy2016,Miszuda2021,Miszuda2022,Handler2020,Fuller2022,Bellinger2024,Henneco2024,Wagg2024}.

In the study of active, close binary systems, processes like mass transfer and/or rotation play a crucial role in shaping the evolution of both components \citep{Sana2012,deMink2013,Henneco2024}. Mass transfer can lead to the expansion or contraction of stellar radii, as well as changes in the orbital period and the dynamical stability of the system \citep[e.g.,][]{Podsiadlowski1992,vanRensbergen2008,Renzo2019}. Additionally, rotation can act as a limiting factor, curbing the donor’s accretion and preventing further mass growth, which may ultimately lead to mass loss from the system \citep[e.g.,][]{Packet1981}. 
Such processes can have far-reaching implications, e.g. in the formation of double neutron star or black hole systems \citep{Vigna-Gomez2020}, exotic stripped star binaries \citep{Laplace2020,Gotberg2020} and production of certain types of supernovae \citep{Renzo2019}.

$\beta$\, Cephei stars in binary systems provide the valuable opportunity to scrutinize evolutionary models for the progenitors of high energy supernovae and gravitational wave events \citep{Tauris2023}. These stars, with masses typically between 8\,\Msun\ and 18\,\Msun, display low-order p- and g-mode pulsations driven by the $\kappa$-mechanism, which is sensitive to opacity variations in iron-group elements within their interiors \citep{Moskalik1992}. In the context of binary systems, eclipsing geometries allow for precise constraints on fundamental stellar parameters, such as masses and radii, which can be directly compared with pulsational and evolutionary models.

Pulsations provide a unique window into stellar interiors and evolutionary processes. As shown by \cite{Miszuda2021,Miszuda2022,Miszuda2024}, variations in pulsation frequencies and mode stability can be used to trace changes induced by mass transfer in binary systems, offering insights into the efficiency of mass accretion and constraining mass transfer rates. These studies highlight how pulsations adapt to altered internal structures, such as changes in density profiles and chemical abundances. Most recently, \cite{Wagg2024} demonstrated that the internal effects of rejuvenation — where accreted material extends the evolutionary lifetime of a star — manifests in specific pulsational signatures, such as modified mode frequencies and amplitudes, enabling constraints on the extent of internal mixing and redistribution of angular momentum. These studies underscore the synergy between asteroseismology and binary evolution theory in unravelling the complexities of stellar interiors.

The V1216\,Sco system (HD\,326440, TIC\,247315421) was first identified as an eclipsing binary of the Algol type by \cite{Otero2004} using ASAS-3 \citep{Pojmanski2002} and Hipparcos \citep{Perryman1997} data, with orbital period of $P_{\rm orb} = 3.9207$\,days and a primary component classified as a B\,0.5\,V star \citep{Kennedy1996}. More recently, \citet{EzeandHandler2024b} detected $\beta$\,Cep-type pulsations in the system’s light curve, utilizing observations from the \TESS space mission. 
In this paper, we present a comprehensive study of the V1216\,Sco system, incorporating both photometric and spectroscopic observations. Section\,\ref{sec:observational_data} provides an overview of the observational data. In Section\,\ref{sec:modelling:spectroscopy}, we detail the spectral analysis, including radial velocity measurements, spectral disentangling, and the determination of atmospheric parameters for both components. The orbital modelling procedure is described in Section\,\ref{sec:modelling:orbital_modelling}, followed by an analysis of the system’s oscillatory behaviour in Section\,\ref{sec:pulsations}. Sections\,\ref{sec:modelling:evolution_modelling} and \ref{sec:puls_modelling} focus on evolutionary and pulsational modelling of the binary system and its components. Finally, the conclusions are summarized in Section\,\ref{sec:conclusions}, while Appendix\,\ref{sec:corner_plots} contains corner plots related to the solutions for radial velocity data, photometric modelling, and the derived absolute parameters from the orbital analysis.

\section{Observational data} 
\label{sec:observational_data}

In this paper we used all observational \TESS \citep[Transiting Exoplanet Survey Satellite,][]{Ricker2015} data that are available at the time of a paper writing. In addition, we supplement them with spectroscopic observations from SALT HRS \citep[South African Large Telescope High Resolution \'echelle Spectrograph,][]{Bramalletal2010}. Below, we discuss the data sets individually.

\subsection{\TESS photometry} 
\label{sec:observational_data:TESS_photometry}

The \system\ was observed by the \TESS space telescope in sectors 12, 39 and 66, with individual image cadences of 30 minutes, 10 minutes, and 2 minutes, respectively. Following these observations, the next data collection is scheduled for June 2025.

To obtain the light curve of the \system, we used \textsc{lightkurve}\footnote{\url{https://docs.lightkurve.org/}} \citep{lightkurve2018}, a python package for \textit{Kepler} and \textit{TESS} data analysis. We extracted $20 \times 20$ pixel cutouts of the original FFI images centred on the target using the \textsc{TESScut} \citep{TESScut} tool. Next the fluxes were extracted from the given cutouts using a 4$\sigma$ threshold with correction for the background signal. We rejected all the data points for which the quality flag value was not zero, indicating that something is wrong with the given flux measurement. 
Such collected light curves were divided into two parts spaced by gaps in each sector, normalised using a quadratic fit to the out-of-eclipse data and sewed back together. 
The final light curve consists of 12\,548 observational points in total, 852 for sector 12, 2\,991 for sector 39 and 8\,705 for sector 66.

\begin{figure*}[htbp]
    \centering
    \includegraphics[width=0.9\textwidth]{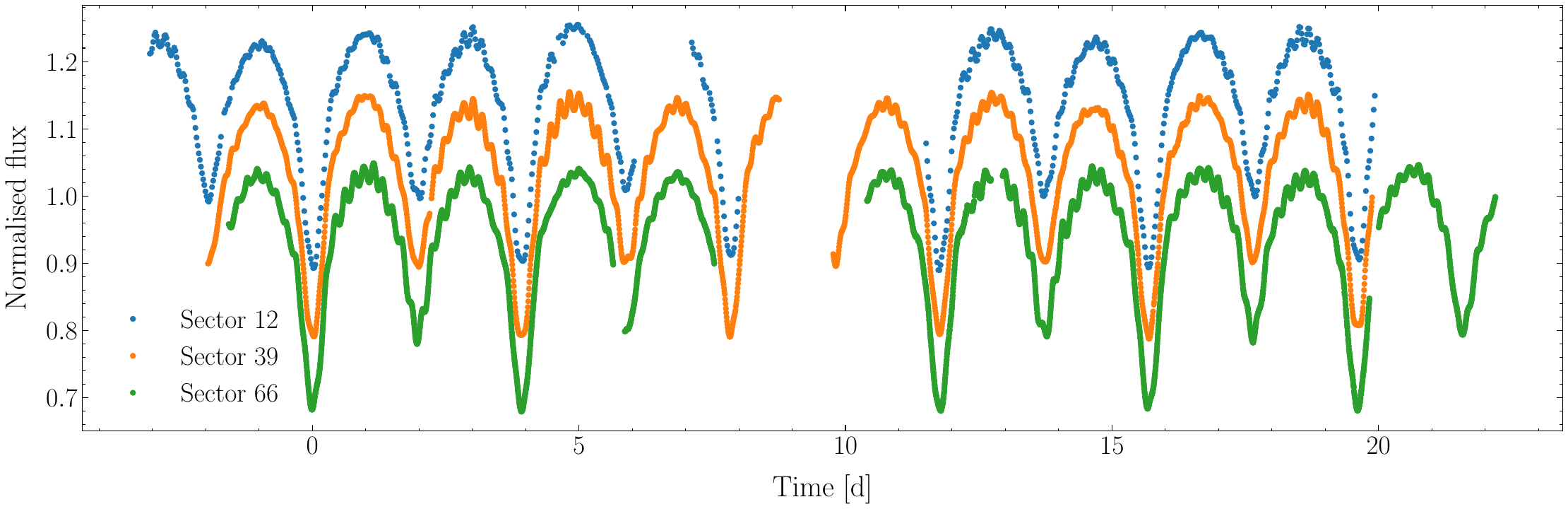}
    \caption{The comparison of the \TESS light curves. The data from each sector was transformed into a common timing by subtracting a BJD value corresponding to the first primary eclipse}
    \label{fig:lc_visualisation}
\end{figure*}

Figure \ref{fig:lc_visualisation} shows a comparison between the data from different sectors. Each subset was shifted such that the first occurrence of superior conjunction occurs at time 0. For clarity, each sector is vertically shifted with a constant offset.

\subsection{Spectroscopy} 
\label{sec:observational_data:spectroscopy}

We conducted a spectroscopic follow-up of V1216\,Sco.  Nine spectra of the target were obtained using the low resolution mode ($R = 14\,000$) of the SALT HRS in 2022.
The SALT HRS is a dual beam (3700-5500\,\AA and 5500-8900\,\AA), fibre-fed, single-object, cross-dispersed \'echelle spectrograph.
The spectra were obtained with an exposure time of 792.8\,s, aiming at a signal-to-noise ratio of 100 at the wavelength of $4567$\AA\, under clear sky conditions. The spectra were reduced using the SALT pipeline \citep{Kniazevetal2016, Kniazevetal2017} and normalized using \textsc{IRAF/PYRAF} tasks \citep{Tody1986}.
For the spectral analysis, we used the blue arm data, i.e., 370-550\,nm.

\section{Spectral analysis}
\label{sec:modelling:spectroscopy}

The orbital variability and atmospheric properties of the target were modelled through a multi-step process. First, we extracted the radial velocities (RVs) from the spectra and independently fitted the RV curves to obtain baseline orbital parameters. Next, we disentangled the spectra, and estimated the atmospheric properties of each component based on the disentangled spectra. 

To extract the RVs, we used a one-dimensional cross-correlation method \citep{TonryandDavis1979, MazehandZucker1994} in which the observed spectrum is cross-correlated with a synthetic template spectrum. The correlation maximum, which is calculated as a function of the shift between the correlated spectra, gives the relative RV of the observed spectrum \citep{TonryandDavis1979, MazehandZucker1994}. Since the system's spectral type is given as B by \cite{Nesterov1995} and our preliminary spectral investigation gives an approximate $T_{\rm eff}$ range between 16\,000\,K and 24\,000\,K, the template synthetic spectrum was generated with an ATLAS model atmosphere \citep{CastelliandKurucz2003} using the parameters of $T_{\rm eff} =18\,000$\,K, $\log g={3.5}$, metallicity $\rm [Fe/H] = 0$ and an isotope line list \citep[hotiso.lst:][]{Gray1999} with the \textsc{spectrum} code \citep{Gray1999}. The lines were broadened with $v\,\sin\,i$ of $70\,\rm km\,s^{-1}$, considering a slightly lower value than the average $v\,\sin\,i$ values for B type stars \citep{Cox2000}, for better radial velocity estimation. 
The RV extraction was performed using the \textsc{iraf/pyraf} {\it fxcor} task \citep{Tody1986, PyRAF2012} using the silicon triplet (Si\,III) lines. The results are given in Table \ref{tab:RVs}. We note that the error estimates on the radial velocities supplied by {\it fxcor} did not appear to be reliable. Therefore we conservatively adopted errors of $\pm15$\,km\,s$^{-1}$ when the lines of the two components were separated, and $\pm30$\,km\,s$^{-1}$ when they were not.

We fitted the orbital elements using the \textsc{rvfit} code \citep{Iglesiasetal2015}. \textsc{rvfit} uses adaptive simulated annealing minimization algorithm for fitting binary and exoplanet radial velocities and has the capacity of fast convergence to a global minimum without the need to provide initial parameter estimates \citep{Iglesiasetal2015}. However, owing to prior information about the orbital period and eccentricity (consistent with zero) of V1216 Sco in the literature \citep[e.g.,][]{EzeandHandler2024b}, we fixed those values during the fitting procedure. The orbital parameters obtained with the \textsc{rvfit} code are shown in Table \ref{tab:preliminary_orbital_parameters}. Those values were used as input priors for the detailed orbital and light curve modelling using the Wilson-Devinney code (see Section\,\ref{sec:modelling:orbital_modelling} for more details).

\begin{table}
    \caption{Radial velocity measurements of V1216\,Sco.}
    \label{tab:RVs}
    \centering 
    \begin{tabular}{crr} 
    \hline\hline 
        TBJD                    & Primary               & Secondary          \T\\ 
        \mbox{[BJD - 2\,457\,000\,d]}  & [$\rm km\,s^{-1}$]    & [$\rm km\,s^{-1}$]  \B\\
        \hline 
        2722.62425 &  -36.6 $\pm$ 30.0 &  -36.6 $\pm$ 30.0 \T\\
        2723.61696 & -132.5 $\pm$ 15.0 &  178.1 $\pm$ 15.0 \\
        2737.31743 &   41.3 $\pm$ 15.0 & -253.8 $\pm$ 15.0 \\
        2738.55639 &  -43.8 $\pm$ 30.0 &  -43.8 $\pm$ 30.0 \\
        2739.54656 & -125.7 $\pm$ 15.0 &  238.0 $\pm$ 15.0 \\
        2740.54656 &  -53.1 $\pm$ 30.0 &  -53.1 $\pm$ 30.0 \\
        2778.45376 & -112.9 $\pm$ 15.0 &  161.9 $\pm$ 15.0 \\
        2797.40668 &  -33.7 $\pm$ 30.0 &  -33.7 $\pm$ 30.0 \\
        2808.37579 &   43.1 $\pm$ 15.0 & -254.7 $\pm$ 15.0 \B\\
    \hline 
    \end{tabular}
\end{table}

\begin{table}
\caption{Preliminary orbital parameters of V1216\,Sco obtained with \textsc{rvfit} code.}
\label{tab:preliminary_orbital_parameters}
\centering 
\setlength{\tabcolsep}{15pt} 
\begin{tabular}{lc} 
\hline\hline 
    Parameter   & Value    \T\B\\ 
    \hline 
    \multicolumn{2}{c}{Adjusted Quantities \T\B}\\
    \hline
    $P_{\rm orb}$\hspace{20pt} [d]		&$3.9213^{\dagger}$ \T\\
    $T_{0,RV}$\hspace{15pt} [TBJD]		&1\,632.73 $\pm$ 0.03\\
    $e$			&$0^{\dagger}$\\
    $\omega$\hspace{29pt} [deg]		&$90^{\dagger}$\\
    $V_{\gamma}$\hspace{26pt} [$ \rm km\, s^{-1}$]	&-36.33 $\pm$ 4.43\\
    $K_1$\hspace{25pt} [$ \rm km\, s^{-1}$]		&88.43 $\pm$ 7.21\\
    $K_2$\hspace{25pt} [$ \rm km\, s^{-1}$]		&237.71 $\pm$ 7.04 \B\\
    \hline
    \multicolumn{2}{c}{Derived Quantities \T\B}\\
    \hline
    $M_1\sin ^3i$\hspace{2pt} [$\rm M_\odot$]	&10.273 $\pm$ 0.888\T\\
    $M_2\sin ^3i$\hspace{2pt} [$\rm M_\odot$]	&3.822 $\pm$ 0.511\\
    $q = M_2/M_1$		&0.372$\pm$ 0.032\\
    $a_1\sin i$\hspace{10pt} [$10^6$ km]	&4.77 $\pm$ 0.39\\
    $a_2\sin i$\hspace{10pt} [$10^6$ km]	&12.82 $\pm$ 0.38\\
    $a  \sin i$\hspace{14pt} [$10^6$ km]	&17.59 $\pm$ 0.54 \B\\
    \hline
\hline 
\end{tabular}
\tablefoot{$^\dagger$ Fixed value}
\end{table}

Binary light curve modelling is a high-dimensional and highly degenerate optimisation problem. Thus, using additional information to constrain or fix parameters is extremely useful to the modelling procedure. Two of the most useful parameters to constrain are the effective temperatures of the two components. To obtain estimates of the primary (the hotter star) effective temperature, we performed spectroscopic disentangling of the composite spectra of V1216\,Sco.
The disentangling was performed using the \textsc{fdbinary} program \citep{Ilijicetal2004}, a C-based code designed for Fourier spectral disentangling of composite spectra that include fluxes from two or three stellar components. For the disentangling, some parameters need to be fixed, such as the flux ratio of the binary components. We estimated the flux ratio to be 1.15 by analysing the double-lined spectra of the system and measuring the equivalent widths of lines from both binary components. In addition to the flux ratio, the orbital parameters obtained from the radial velocity analysis were fixed during the disentangling analysis, except for $P_{\rm orb}$ and $T_{0,RV}$. Since the hydrogen lines are very sensitive to the values of $T_{\rm eff}$ and $\log g$ in B-type stars, we mostly focused on the $H_{\beta}$ line. We also used He\,I lines (e.g. at 4387.9\,\AA) to estimate $v\,\sin\,i$ values of the binary components. 

\begin{figure*}[thbp]
    \centering
    \includegraphics[width=0.9\textwidth]{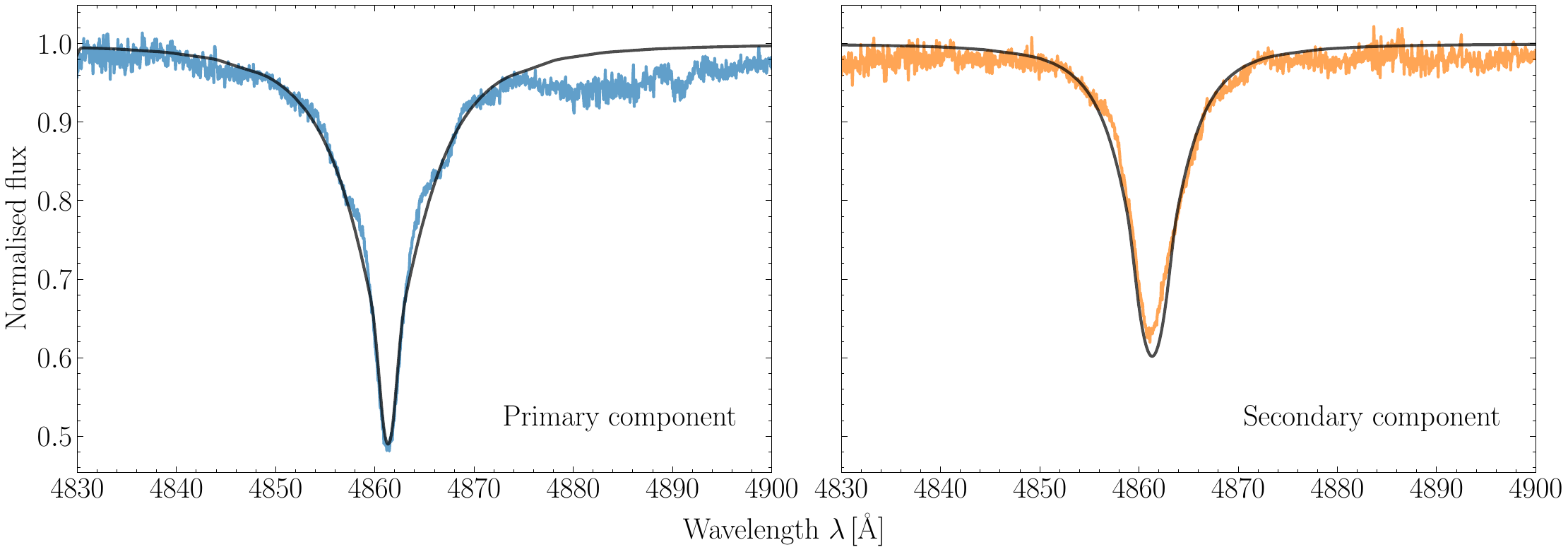}
    \caption{The disentangled spectra of V1216\,Sco showing the $H_{\beta}$ lines for primary and secondary components (left and right panels, respectively), with the best fitting atmospheric models.}
    \label{fig:spectra_fit}
\end{figure*}

Using the disentangled spectra, we estimated $T_{\rm eff}$, $\log g$ and $v\,\sin\,i$ for each binary component. To determine these parameters, the Non-Local Thermodynamic Equilibrium (NLTE) Tlusty BSTAR2006 grid \citep{HubenyandLanz2017} model atmospheres were used. The values were derived by comparing the observed lines with the synthetic ones and by minimizing the difference between the observed and the synthetic model, as described by \citet{Catanzaroetal2004}. 
The $v\,\sin\,i$ measurements, however, were derived from He\,I lines using a profile-fitting method. This involved comparing limb-darkened, instrumentally broadened models in steps of $1\, \rm km\, s^{-1}$ to the observed He\,I line profiles.
The atmospheric solutions of the target are summarised in Table \ref{tab:spectroscopy_parameters} with $1\sigma$ errors. The best fitting atmospheric models to the observed $H_{\beta}$ lines for both components are shown in Figure \ref{fig:spectra_fit}.  

To investigate the effect of possible mass transfer on the spectra we scanned He and $H_{\alpha}$ lines using the blue and red arm data. The spectra do not show any clear emission and/or absorption lines \citep{Vesper2001} that might suggest ongoing mass transfer in the system. Nonetheless, mass transfer, in certain orbital phases, may create asymmetries in the line profiles rather than strong emission features \citep[e.g.,][]{Vesper1993}. To investigate this possibility, we analysed the residuals obtained by subtracting the model composite of $H_{\alpha}$ lines from the observed line profiles. However, no significant asymmetries were detected, suggesting that any potential mass transfer effects are either minimal or not observable within the sensitivity of our data.

\begin{table}
\caption{Spectral parameters of V1216\,Sco.}
\label{tab:spectroscopy_parameters}
\centering 
\setlength{\tabcolsep}{12pt} 
\begin{tabular}{lcc} 
\hline\hline 
    Parameter   & Primary &  Secondary   \T\B\\ 
    \hline 
    $T_{\rm eff}$ \hspace{13pt}$\rm[K]$            & $22\,000 \pm 1\,000$ & $20\,000 \pm 1\,500$ \T\\
    $\log g$ \hspace{8pt}$\rm[dex]$               & $3.75 \pm 0.20$ & $3.20 \pm 0.20$ \\
    $v\,\sin\,i$ \hspace{2pt}$\rm[km\, s^{-1}]$    & $95 \pm 5$ & $115 \pm 5$ \B\\
    \hline
\hline 
\end{tabular}
\end{table}

\section{Orbital modelling}
\label{sec:modelling:orbital_modelling}

The radial velocities and the \TESS photometry from sectors 12, 39 and 66 of \system\ were analysed using the 2015 version of the Wilson-Devinney (\textsc{WD}) program \citep{Wilson1971,Wilson1979,Wilson1990} supplied with the \TESS band pass\footnote{https://faculty.fiu.edu/\tylde vanhamme/lcdc2015/}.
For the error estimation of the derived parameters we used the Wilson-Devinney \textsc{LC} code, called within our own Markov Chain Monte Carlo (MCMC) framework working under the \textsc{emcee}\footnote{https://emcee.readthedocs.io/en/stable/} package \citep{Foreman_Mackey_2013}, a \textsc{python} implementation of Goodman \& Weare’s Affine Invariant MCMC Ensemble sampler to calculate the posterior probability distribution for the model parameters.

\subsection{Preparation of the photometric data}
The pulsations have relatively strong amplitudes compared to the eclipse depths (see Figure \ref{fig:lc_visualisation}), hence leaving them in the data could impact the parameter estimation. To correct the data for the strongest pulsation signals, we used the Fourier method to identify and subtract the subsequent strongest frequencies \citep[see, e.g.,][]{Miszuda2022}, other than orbital harmonics, whose amplitudes were above 1\,ppm. The subtracted frequencies were $f=$ 5.524, 5.680, 5.572 and 5.412\,\cpd. 
Next, to find an orbital  period ($P_{\rm orb} = 1/f_{\rm orb}$) of the \system, we once again calculated the periodogram for the cleaned data. The highest orbital peak visible in the periodogram ($f=$ 0.508\,\cpd) corresponds to half of the orbital period (since there are two eclipses per orbit), so for an initial guess we took half of that value and simultaneously fitted all previously found frequencies with 50 orbital harmonics, obtaining $P_{\rm orb} = 3.92063$\,d. 

The light curve of the V1216 Sco shows a small O'Connell effect. This phenomenon, often visible in binary light curves, is related to visible asymmetries in the flux brightness at quadratures \citep{Milone1968,Davidge1984}. 
The most common explanation for the O’Connell effect is the presence of cool or hot spots on the surface of one or both stellar components, typically resulting from localized magnetic activity or mass transfer processes. In non-synchronized systems, where the orbital period and rotational period of the stars differ, the asymmetry can also be attributed to irradiation and reflection effects. These occur when one star’s radiation heats specific regions of the other star’s surface, creating flux imbalances that depend on the orbital phase. Non-synchronization can amplify these effects, leading to observable flux variations at quadrature points.
Due to those uncertainties and because there is no unique explanation for the O'Connell effect we corrected the flux by fitting sine and cosine terms, following the \cite{Faigler2011} prescription. 

In the next step, we binned the light curve from each sector into 100 phase bins. To do this, we phase-folded the observations and selected binning intervals such that the distance between each bin point and its corresponding flux - calculated as the median flux within each bin - remained approximately constant across the orbital phase-normalised flux plane. In that way the narrow eclipses were sampled more densely than areas outside of them and the out-of-eclipse variability did not dominate the solution. The phased light curves were later transformed back to the time domain, so they no longer assumed $P_{\rm orb}$ and $T_0$ values used to phase the observations in the first place.

\subsection{Comprehensive modelling of radial velocities and \TESS photometry}
To obtain an orbital model of V1216 Sco system, in a first step we analysed the light curve from Sector 66. We fixed the values of the orbital period and the time of superior conjunction to $P_{\rm orb} = 3.92063$\,d and $T_{0, TESS} = 3099.265$\,BJD, respectively. We used the primary effective temperature $T_{\rm eff, 1} = 22\,000$\,K and the mass ratio $q=0.372$ estimates from the previous spectral analysis (Section\,\ref{sec:modelling:spectroscopy}).
The limb darkening (LD) bolometric coefficients and the \TESS LD coefficients were obtained for the logarithmic LD law from \cite{vanHamme1993} and \cite{Claret2017_TESS_LD}, respectively.

The initial models were calculated with the assumption of detached geometry, 
with fitting for eccentricity, periastron argument, inclination, effective temperature of the secondary, luminosity of the more massive star and potentials of both components. 
We noticed that the radius of the secondary star approaches the critical radius, at which the star fills its Roche lobe. Consequently, we repeated our calculation assuming the secondary star fills its Roche lobe.
This model provided a slightly worse fit, particularly near the secondary eclipse, during contact moments, so we decided to discard the assumption of a semi-detached configuration and revert to a fully detached one.
In both modes, the eccentricity value always reached very small values, $\sim 0.01$, so in the next runs we fixed it to $e=0.0$.

\begin{figure*}[htbp]
    \centering
    \includegraphics[width=0.49\textwidth]{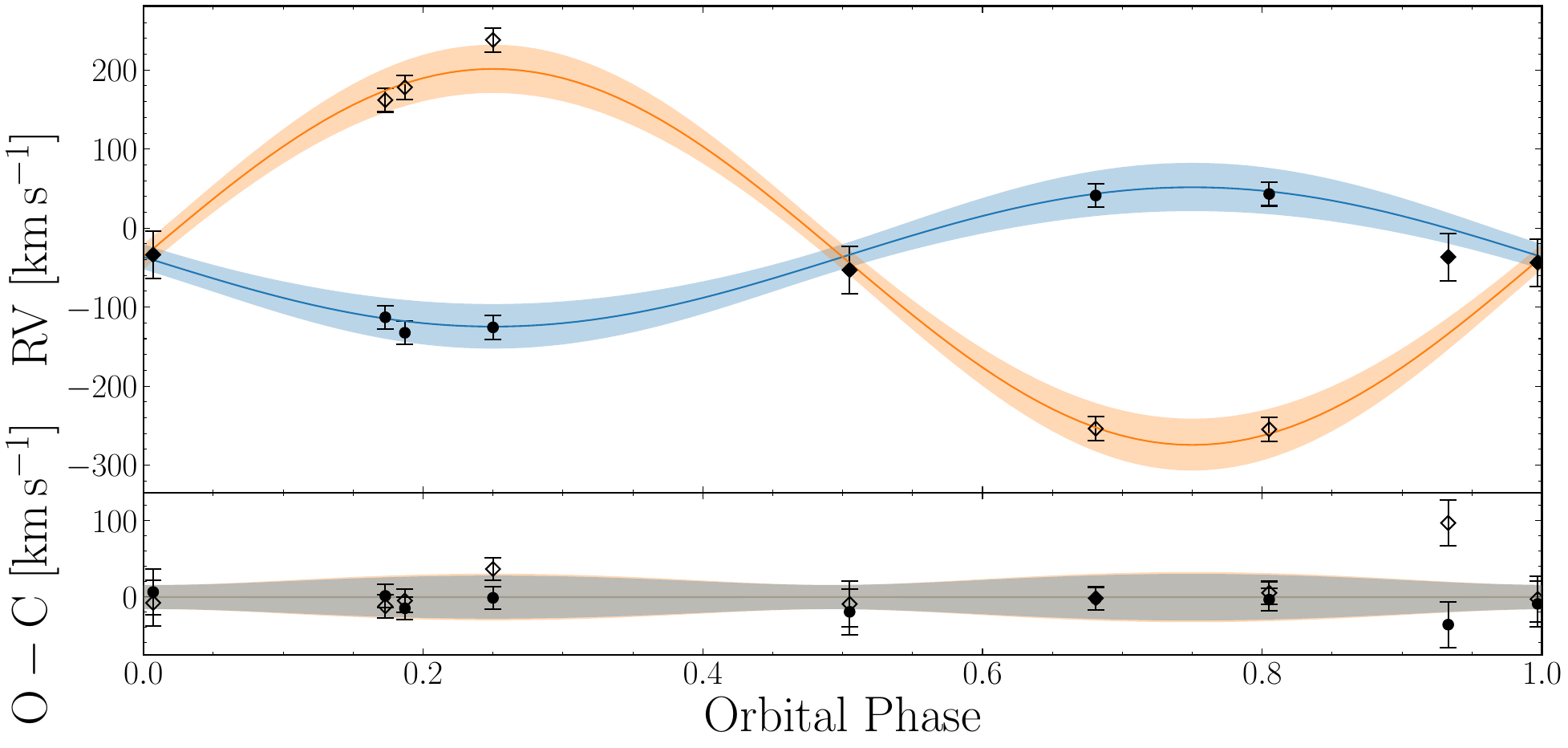} 
    \includegraphics[width=0.48\textwidth]{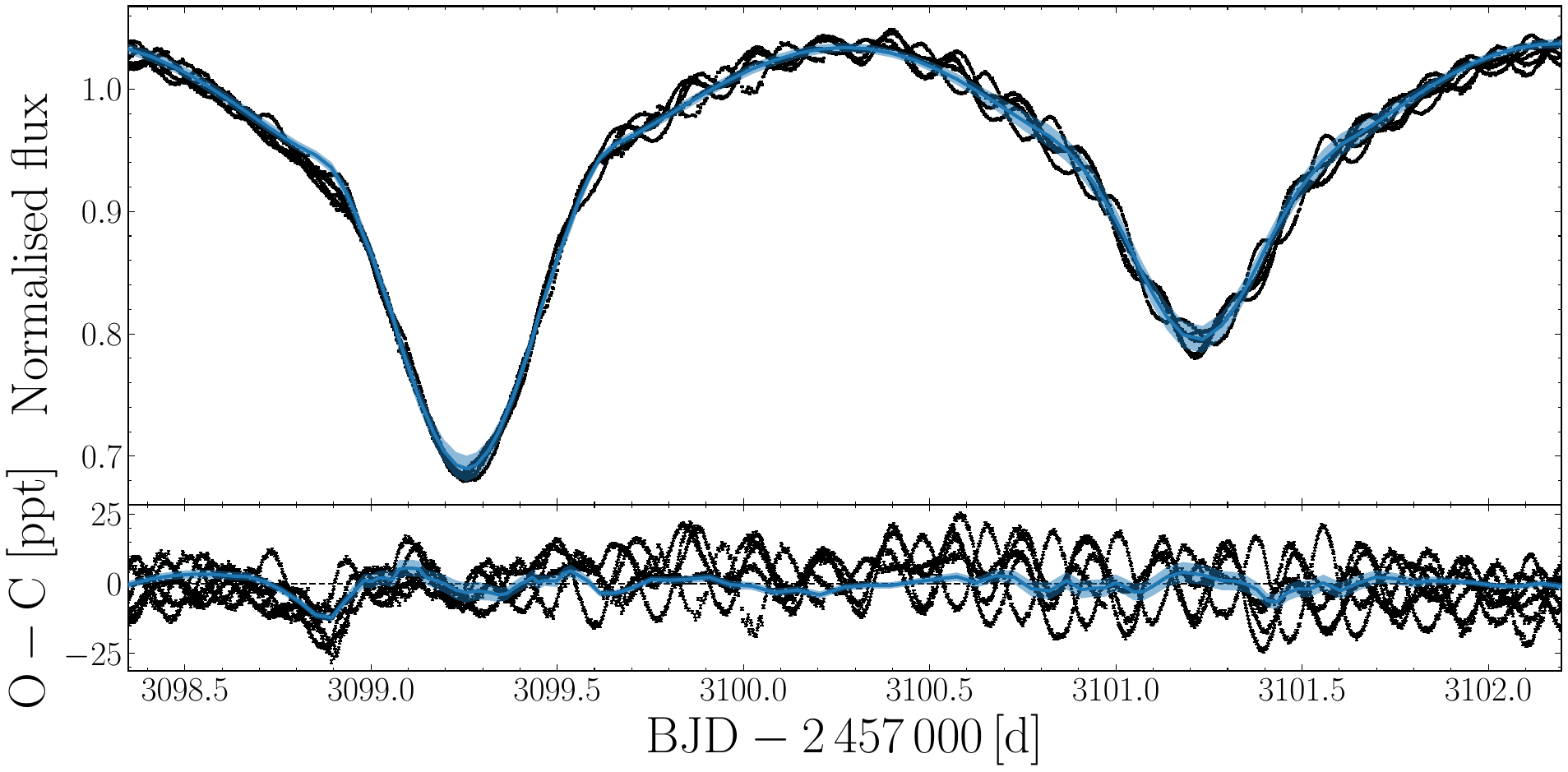} 
    
    \caption{Fitting results of the V1216 Sco binary system. The left panel displays the fits for radial velocities, while the right panel shows the light curve fit. Each fit shows the corresponding residuals (bottom panels), with shaded 3$\sigma$ credible regions indicated based on MCMC sampling.}
    \label{fig:RV-LC_fit}
\end{figure*}

Using the parameter values obtained from previous runs, we refined the spectral parameters by fitting exclusively to the radial velocity measurements. Specifically, we fitted for the time of superior conjunction $T_{0, RV}$, the mass ratio $q \equiv M_2/M_1$, the semi-major axis $a$ and the systemic velocity $V_{\gamma}$. We show the results in Table \ref{tab:WD_spectral_parameters}. We estimated the errors using Markov Chain Monte Carlo (MCMC) methods, employing 20 walkers and 5\,000 iterations, with an initial burn-in phase of 1\,000 iterations. The left panel of Figure \ref{fig:RV-LC_fit} illustrates the best fit to the radial velocity data, and we provide the corner plots from the MCMC sampling in Appendix Figure \ref{fig:RV_corner}.

\begin{table}
\caption{Spectroscopic orbital parameters of V1216\,Sco.}
\label{tab:WD_spectral_parameters}
\centering 
\setlength{\tabcolsep}{20pt} 
\begin{tabular}{lc} 
\hline\hline 
    Parameter   & Value     \T\B\\ 
    \hline 
    $T_{0,RV}$ \hspace{4pt}$\rm [TBJD]$ & $2\,722.8564 ^{+0.0303} _{-0.0303}$ \T\\
    $P_{\rm orb}$ \hspace{8pt}$\rm [d]$ & $3.92063 ^{\dagger}$ \\
    $q \equiv M_2/M_1$ & $0.372 ^{+0.032} _{-0.032}$ \\
    $a$ \hspace{20pt}$\rm [R_{\odot}]$ & $26.42 ^{+0.85} _{-0.84}$ \\
    $V_{\gamma}$ \hspace{14pt}$\rm [km\,s^{-1}]$ & $-36.32 ^{+4.33} _{-4.41}$ \B\\
    \hline
\hline 
\end{tabular}
\tablefoot{The listed parameter values represent the mode of the posterior distributions, with uncertainties provided as the $16^{\rm th}$ and $84^{\rm th}$ percentiles, corresponding to the $1\sigma$ credible intervals. \\
$^\dagger$ Fixed value}
\end{table}

Having refined the radial velocity modelling we once again approached the light curve modelling, this time for each \TESS\ sector. 
One of the most noticeable features in the V1216 Sco light curve is the flux asymmetry observed between the shoulders of the primary eclipses. This characteristic is notably stable, as it persists across sectors that are approximately one year apart. As the secondary component is very close to filling it's Roche lobe, based on preliminary estimates of its radius, we assume that the system is either currently undergoing mass transfer or has just passed that phase. Consequently, the observed flux dip could be attributed to material remnants that block the emergent fluxes at specific orbital phase. To account for this feature during modelling and obtain more reliable radius estimates, we employed Gaussian Processes (GPs) using the \textsc{celerite} package \citep{celerite}, that allow to model astrophysical noise (like spots or other features) without complicating the model with vast number of additional parameters. In the final run of the \textsc{WD} modelling, we fitted for the time of superior conjunction $T_{0, TESS}$ along with inclination $i$, effective temperature of the secondary $T_{\rm eff,2}$, both potentials $\Omega_{1,2}$ and \TESS\ primary's passband luminosity $L_{TESS}$. Similarly to the RVs, the errors were estimated using the Markov Chains with 50 walkers and 5\,000 iterations with a preceding burn-in phase set to 1\,000 iterations. To account for the underestimation of observational errors, during the MCMC run we also sampled for the noise nuisance parameter $\sigma_{\ln f}$ following the description of \cite{Conroy2020}. The MCMC runs were repeated for each sector of the \TESS\ observations, and the common solution was obtained through Monte Carlo propagation of priors. The results are summarised in Table \ref{tab:WD_photometric_parameters} and the best fit for sector 66 is illustrated in the right panel of Figure \ref{fig:RV-LC_fit}. We also provide the corner plots from the MCMC sampling for sector 66 data in Appendix Figure \ref{fig:LC_corner}.

\begin{table*}
\caption{Light curve parameters of V1216\,Sco.}
\label{tab:WD_photometric_parameters}
\centering 
\setlength{\tabcolsep}{6pt} 
\begin{tabular}{lcccccccc} 
\hline\hline 
  & \multicolumn{2}{c}{\textsc{Sector 12}} & \multicolumn{2}{c}{\textsc{Sector 39}} & \multicolumn{2}{c}{\textsc{Sector 66}} & \multicolumn{2}{c}{\textsc{Median values}} \T\\
Parameter & Primary & Secondary & Primary & Secondary & Primary & Secondary & Primary & Secondary \B\\
    \hline 
    $T_{0,TESS}$ \hspace{3pt}$\rm [TBJD]$ & \multicolumn{2}{c}{$1\,632.9490 ^{+0.0012} _{-0.0012}$} & \multicolumn{2}{c}{$2\,366.1024 ^{+0.0012} _{-0.0012}$} & \multicolumn{2}{c}{$3\,099.2569 ^{+0.0010} _{-0.0009}$} & \multicolumn{2}{c}{---} \T\\
    $P_{\rm orb}$ \hspace{18pt}$\rm [d]$ & \multicolumn{2}{c}{$3.92063 ^{\dagger}$} & & & & & \multicolumn{2}{c}{$3.92063 ^{\dagger}$} \\
    $i$ \hspace{32pt}$\rm [\rm deg]$ & \multicolumn{2}{c}{$73.00 ^{+0.20} _{-0.18}$} & \multicolumn{2}{c}{$73.21 ^{+0.14} _{-0.15}$} & \multicolumn{2}{c}{$73.06 ^{+0.13} _{-0.14}$} & \multicolumn{2}{c}{$73.09 ^{+0.18} _{-0.18}$} \\
    $q \equiv M_2/M_1$ & \multicolumn{2}{c}{$0.372 ^{\dagger}$} & & & & & \multicolumn{2}{c}{$0.372 ^{\dagger}$} \\
    $e$ & \multicolumn{2}{c}{$0.0 ^{\dagger}$} & & & & & \multicolumn{2}{c}{$0.0 ^{\dagger}$} \\
    $L_3$ \hspace{25pt}$\rm [\%]$ & \multicolumn{2}{c}{$0.0 ^{\dagger}$} & & & & & \multicolumn{2}{c}{$0.0 ^{\dagger}$} \\
    $T_{\rm eff}$ \hspace{21pt}$\rm [K]$ & $22\,000^{\dagger}$ & $17\,235 ^{+243} _{-255}$ & $22\,000^{\dagger}$ & $17\,637 ^{+242} _{-241}$ & $22\,000^{\dagger}$ & $17\,208 ^{+216} _{-219}$ & $22\,000^{\dagger}$ & $17\,336 ^{+341} _{-296}$\\
    $\Omega$ & $3.585 ^{+0.070} _{-0.052}$ & $2.592 ^{+0.003} _{-0.004}$ & $3.587 ^{+0.043} _{-0.043}$ & $2.582 ^{+0.002} _{-0.002}$ & $3.570 ^{+0.032} _{-0.044}$ & $2.596 ^{+0.002} _{-0.003}$ & $3.579 ^{+0.048} _{-0.046}$ & $2.591 ^{+0.005} _{-0.010}$ \\
    $A$ & $1.0^{\dagger}$ & $1.0^{\dagger}$ & & & & & $1.0^{\dagger}$ & $1.0^{\dagger}$  \\
    $g$ & $1.0^{\dagger}$ & $1.0^{\dagger}$ & & & & & $1.0^{\dagger}$ & $1.0^{\dagger}$  \\
    $L_{TESS}$ & $7.885^{+0.146} _{-0.160}$ & --- & $7.731^{+0.131} _{-0.118}$ & --- & $8.166^{+0.127} _{-0.103}$ & --- & \multicolumn{2}{c}{---}\\
    $\sigma_{\ln f}$ & \multicolumn{2}{c}{$-5.219^{+0.082} _{-0.078}$} & \multicolumn{2}{c}{$-5.244^{+0.086} _{-0.083}$} & \multicolumn{2}{c}{$-5.450^{+0.085} _{-0.084}$} & \multicolumn{2}{c}{---}\\
    $x$ & $0.3247^{\dagger}$ & $0.4077^{\dagger}$ & & & & & $0.3247^{\dagger}$ & $0.4077^{\dagger}$ \\
    $y$ & $0.2129^{\dagger}$ & $0.2778^{\dagger}$ & & & & & $0.2129^{\dagger}$ & $0.2778^{\dagger}$ \\
    $x_{\rm bol}$ & $0.758^{\dagger}$ & $0.704^{\dagger}$ & & & & & $0.758^{\dagger}$ & $0.704^{\dagger}$ \\
    $y_{\rm bol}$ & $0.112^{\dagger}$ & $0.068^{\dagger}$ & & & & & $0.112^{\dagger}$ & $0.068^{\dagger}$ \B\\
    \hline
\hline 
\end{tabular}
\tablefoot{The listed parameter values represent the mode of the posterior distributions, with uncertainties provided as the $16^{\rm th}$ and $84^{\rm th}$ percentiles, corresponding to the $1\sigma$ credible intervals. The bolometric LD coefficients, $x_{\rm bol}$ and $y_{\rm bol}$, are from \cite{vanHamme1993}, while $x$ and $y$, the \TESS LD coefficients were taken from \cite{Claret2017_TESS_LD}.\\
$^\dagger$ Fixed value for all sectors}
\end{table*}

Although the overall fit is satisfactory, the \TESS\ residual light curve reveals some systematic differences between the best model and the observations (see the bottom right panel of Figure \ref{fig:RV-LC_fit}). The largest scatter, on the order of 10 ppt, can be seen around the times of first contact during the primary eclipse, most likely associated with presence of gaseous remnants. We remind that this feature was accounted for during the fitting with GPs. However, to ensure that the GPs do not overfit an incorrect model to the data, the sensitivity of the GPs was set to be relatively low (i.e. $\sigma = -5$). 
Just as selecting the right bottle of wine requires careful consideration to avoid overwhelming the senses, fine-tuning the GPs demands a similar level of precision to ensure the model is not overfitted to any minor fluctuation.
While this low setting couldn’t entirely remove the observed flux dip, it was a necessary compromise to avoid overfitting to any minor fluctuation. Additionally, since the model fits the right wing of the primary eclipse perfectly (while the left wing does not), the risk of deriving incorrect radii is negligible.

Using the results from the radial velocities and the median results from photometric light curve modelling, we derived the absolute parameters of the V1216\, Sco components, which are summarised in Table \ref{tab:WD_absolute_parameters} and can be seen in Appendix Figure \ref{fig:MR_corner}. V1216\, Sco is a high-mass binary system, containing $11.72\,\rm M_{\odot}$ and $4.35\,\rm M_{\odot}$ stars. We note that, despite large difference in their masses, both components have similar radii -- $8.4\,\rm R_{\odot}$ for the primary and $8.1\,\rm R_{\odot}$ for the secondary. While the radius of the primary is typical for a $\sim 12\,\rm M_{\odot}$ star halfway through the main sequence, the secondary radius is far to large to be explained as a product of a $\sim 4\,\rm M_{\odot}$ single star evolution that have not reached a giant phase.
Therefore, we conclude that the V1216\,Sco is a mass transfer or post-mass transfer system, with the secondary component very close to the critical radius.

\begin{table}
\caption{Absolute parameters of V1216\,Sco.}
\label{tab:WD_absolute_parameters}
\centering 
\setlength{\tabcolsep}{15pt} 
\begin{tabular}{lcc} 
\hline\hline 
    Parameter   & Primary   & Secondary     \T\B\\ 
    \hline 
    Mass  \hspace{10pt} $\rm[M_{\odot}]$ & $11.720^{+1.210} _{-1.130}$ & $4.344 ^{+0.657} _{-0.637}$ \T\\
    Radius \hspace{3pt} $\rm[R_{\odot}]$ & $8.426^{+0.139} _{-0.105}$  & $8.058 ^{+0.024} _{-0.017}$ \\
    $\log L/ \rm L_{\odot}$              & $4.174^{+0.080} _{-0.080}$  & $3.970 ^{+0.130} _{-0.130}$ \B\\
    \hline
\hline 
\end{tabular}
\tablefoot{The listed parameter values represent the mode of the posterior distributions, with uncertainties provided as the $16^{\rm th}$ and $84^{\rm th}$ percentiles, corresponding to the $1\sigma$ credible intervals.}
\end{table}

\section{Oscillation data analysis}
\label{sec:pulsations}

To extract the pulsational behaviour of the system, we performed Fourier analysis on the light curves corrected for the binary orbit. For this, we used the residuals from all three \TESS sectors, obtained by subtracting the \WD\ model from the data.

We calculated the amplitude spectra using a discrete Fourier transform \citep{Deeming1975,Kurtz1985}. Whereas the amplitude spectra of the light curves from the different sectors are qualitatively similar, they do show some quantitative differences. The dominant oscillation frequencies are similar, but the amplitudes of those signals differ somewhat from sector to sector. This may indicate intrinsic changes in the amplitudes, but could also be due to beating of multiple frequencies that are unresolved within the individual sectors' data. As it is not possible to separate these two hypotheses with the available data set, we determined the frequency content from each sector of data separately. Apart from that the sampling rate of the light curves is different in each sector. However, all the variability occurs at frequencies below the Nyquist frequency for the Sector 12 data set.

We followed the standard prewhitening procedure, computing periodograms up to the Nyquist frequency for the Sector 12 data, $f_{\rm N} \sim 24$\,\cpd. We established a signal-to-noise ratio limit of $\rm S/N=5$ as a threshold for significant frequencies \cite[see][]{Baran2021}. 
The Fourier noise level was determined as an average amplitude value within a 5\,\cpd\ sliding window.

Despite subtracting the orbital model from the data, we noticed the presence of integer multiples of the orbital frequency in the periodograms. These signals can arise from trends in the residuals (see right-bottom panel of Figure \ref{fig:RV-LC_fit}) and difficulties encountered during the light curve normalisation procedure, as the depths of the eclipses that can vary slightly in each sector. To address these signals, we further corrected the data for 100 orbital harmonics. 

With the iterative approach, we searched for frequencies with the highest amplitudes that satisfied the condition $S/N \ge 5$. We identified eight peaks that occupy the region from $\sim 5.4$\,\cpd\ to $\sim 11.1$\,\cpd. The three signals above 10\,\cpd\ are consistent with combination frequencies of the two strongest pulsations. The eight frequencies with their amplitudes and respective formal error estimates \citep[following][]{1999DSSN...13...28M} are presented in Table \ref{tab:frequencies}. 

\begin{figure*}[tbhp]
    \centering
    \includegraphics[width=0.9\textwidth]{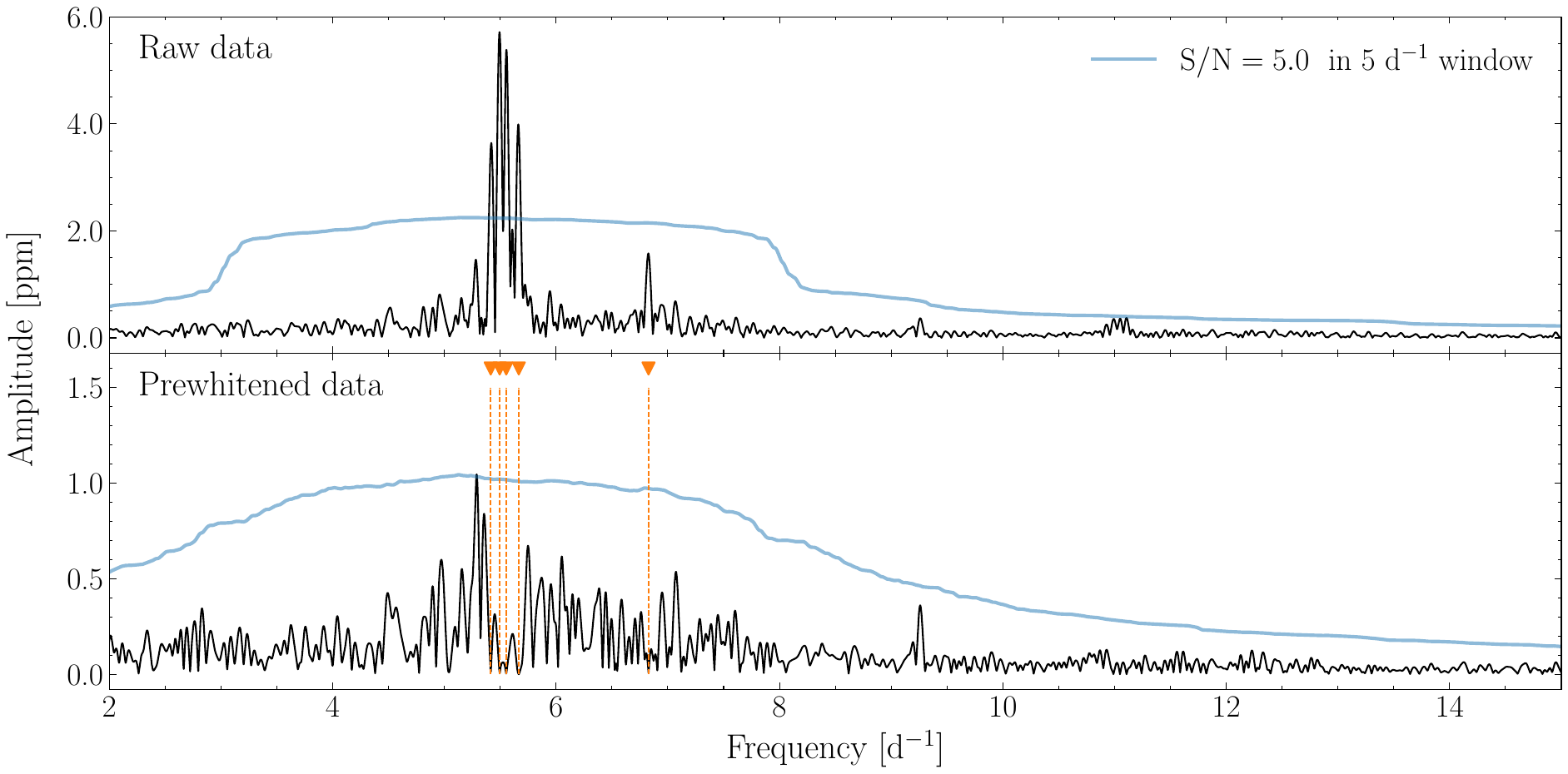} 
    \caption{Periodograms calculated from the \TESS sector 66 data, corrected for the binary orbit, are presented along with their corresponding signal-to-noise (S/N) curves. The top panel shows the periodogram for the original data, while the bottom panel presents the periodogram after prewhitening, highlighting the five independent frequencies. These frequencies are marked by vertical lines for clarity.}
    \label{fig:periodogram}
\end{figure*}

The periodograms for the original and prewhitened residuals from orbital modelling are shown in Figure \ref{fig:periodogram}, in the top and bottom panels, respectively, along with the $S/N = 5$ curves. The detected signals are marked with orange dashed lines. As one can see, unresolved signals remain in the data, specifically at $f \approx 5.3$\,\cpd\ and $f \approx 9.3$\,\cpd; however, our ability to detect these signals is currently limited by the length of the individual \TESS data sets. Two consecutive sectors of observation would likely resolve the pulsation spectrum fully.

\begin{table*}
\centering
\caption{Pulsation frequencies in the \TESS\ light curves of V1216\,Sco.}
\setlength{\tabcolsep}{5pt} 
\begin{tabular}{lccrccrccr}
\hline
\hline
& \multicolumn{3}{c}{\textsc{Sector 12}} & \multicolumn{3}{c}{\textsc{Sector 39}} &  \multicolumn{3}{c}{\textsc{Sector 66}} \T\\
ID & Frequency & Amplitude &  S/N & Frequency & Amplitude &  S/N & Frequency & Amplitude &  S/N \\
& [d$^{-1}$] & [ppt] & & [d$^{-1}$] & [ppt] & & [d$^{-1}$] & [ppt] & \B\\
\hline
$f_1$ & 5.5533 $\pm$ 0.0005 & 5.49 $\pm$ 0.12 & 22.4 & 5.5507 $\pm$ 0.0005 & 4.84 $\pm$ 0.10 & 20.8 & 5.5536 $\pm$ 0.0005 & 4.30 $\pm$ 0.09 & 21.1 \\
$f_2$ & 5.4965 $\pm$ 0.0011 & 2.69 $\pm$ 0.12 & 11.0 & 5.4947 $\pm$ 0.0006 & 4.02 $\pm$ 0.10 & 17.3 & 5.4950 $\pm$ 0.0005 & 4.52 $\pm$ 0.09 & 22.1 \\
$f_3$ & 5.6709 $\pm$ 0.0009 & 3.38 $\pm$ 0.12 & 13.9 & 5.6628 $\pm$ 0.0009 & 2.93 $\pm$ 0.10 & 12.5 & 5.6661 $\pm$ 0.0006 & 3.50 $\pm$ 0.09 & 17.3 \\
$f_4$ & 5.4182 $\pm$ 0.0011 & 2.71 $\pm$ 0.12 & 11.0 & 5.4121 $\pm$ 0.0010 & 2.54 $\pm$ 0.10 & 11.0 & 5.4149 $\pm$ 0.0007 & 2.89 $\pm$ 0.09 & 14.1 \\
$f_5$ & 6.8278 $\pm$ 0.0015 & 2.00 $\pm$ 0.12 & 9.2 & 6.8306 $\pm$ 0.0011 & 2.35 $\pm$ 0.10 & 10.7 & 6.8275 $\pm$ 0.0014 & 1.50 $\pm$ 0.09 & 7.8 \\
$2f_2$ & 10.9929 & 0.28 $\pm$ 0.12  & 3.8  & 10.9894 & 0.29 $\pm$ 0.10  & 5.6  & 10.9900 & 0.40 $\pm$ 0.09  & 7.0 \\
$f_1+f_2$ & 11.0497 & 0.30 $\pm$ 0.12  & 4.2  & 11.0455 & 0.21 $\pm$ 0.10  & 4.1  & 11.0486 & 0.23 $\pm$ 0.09  & 4.2 \\
$2f_1$ & 11.1065 & 0.30 $\pm$ 0.12  & 4.2  & 11.1015 & 0.39 $\pm$ 0.10  & 7.8  & 11.1072 & 0.31 $\pm$ 0.09  & 5.6 \\
\hline
\end{tabular}
\label{tab:frequencies}
\end{table*}

\section{Modelling of the binary evolution} 
\label{sec:modelling:evolution_modelling}

Because our light curve modelling showed that the secondary radius is way too large for its mass to be explained as a product of a single-star evolution, we suspect that \system\ is a product of a system that has passed through a mass transfer phase.
Therefore, \system\ presents a challenge in the context of modelling the evolution of its components. 
To model the system not as two isolated stars, but as an effect of the binary interactions in the past, we used the \textsc{mesa} code \citep[Modules for Experiments in Stellar Astrophysics,][version 23.05.1]{Paxton2011, Paxton2013, Paxton2015, Paxton2018, Paxton2019, Jermyn2023}, with the \textsc{mesa-binary} module. The information on \textsc{mesa} input physics can be found in Appendix\,\ref{sec:mesa_physics}.

In order to reproduce the evolution of the system, we built an extensive grid of evolutionary models. We constructed a set of parameters to be varied, i.e. initial orbital period, initial masses of the components, metallicity, overshooting from the convective core, mixing-length theory scaling coefficient and a fraction of mass lost during mass transfer (MT), with the ranges as given in Table \ref{tab:MESA_parameter_ranges}. From this set we constructed 5\,000 vectors of initial parameters with the assumed numerical accuracy (see last column of Table \ref{tab:MESA_parameter_ranges}) and calculated the evolutionary tracks. The parameters that characterise the binary evolutionary tracks were chosen randomly from uniform distributions within the given ranges. All of these parameters were chosen independently of others, except for the masses.
The masses of the individual stars were varied between $3\, \rm M_{\odot}$ and $15\, \rm M_{\odot}$. However, based on previous mass estimates we assumed that the total initial mass of the system should be no less than $16\, \rm M_{\odot}$. We tested the rate of mass transfer loss from the system as a fraction of mass that is lost from the vicinity of the accretor in a form of a fast wind during MT, in the range from $\beta = 0.0$ to $\beta=0.3$. The initial orbital periods of our models span 1 to 7 days.

In our evolutionary computations, we used the AGSS09 \citep{Asplund2009} initial chemical composition of the stellar matter and the OPAL opacity tables. These tables were supplemented with data from \cite{Ferguson2005} for lower temperatures. For higher temperatures, hydrogen-poor or metals-rich conditions we used C and O enhanced tables. To determine the chemical composition of stars we varied the value of metallicity $Z$, from $10^{-3}$ to $3 \times 10^{-2}$ and used the following scaling relations:
\begin{eqnarray}
   Y &=& Y_p + \left( \frac{Y_{\rm protosolar} - Y_p}{Z_{\rm protosolar}} \right) Z, \\
  X &=& 1 - Z - Y, 
\end{eqnarray}
with primordial helium abundance $Y_p=0.249$ \citep{PlanckCollaboration2016}, and $Y_{\rm protosolar}=0.2703$, $Z_{\rm protosolar}=0.0142$ protosolar abundances \citep{Asplund2009}.

We adopted the Ledoux criterion for convective instability, combined with turbulent convection theory based on the \cite{Kuhfuss1986} model, that at long timescales reduces to the \cite{Cox1968} model. We used a mixing-length scaling factor, $\alpha_{\rm MLT}$, ranging from 0.0 to 1.5. To correctly define the convective boundaries, we employed the convective premixing algorithm, which expands the boundaries until $\nabla_{\rm rad} = \nabla_{\rm ad}$ on the convective side. This algorithm also allows for adjusting the composition gradient to achieve convective neutrality, thereby establishing semi-convective regions. In \textsc{mesa}, these follow the formalism of \cite{Langer1985}, which we modify using the free parameter $\alpha_{\rm SC} = 0.01$. Additionally, we incorporated overshooting mixing beyond the formal convective boundaries, using an overshooting parameter, $f_{\rm ov}$, to account for turbulent motions extending into the radiative zone. We applied an exponential overshooting scheme \citep{Herwig2000} for all burning zones, including both the core and shell. The value of $f_{\rm ov}$ above the convective boundaries was set as a free parameter, spanning from $10^{-3}$ to $4 \times 10^{-2}$, while below these zones we adopted a fixed value of $f_{\rm ov} = 0.01$. In all cases, we set $f_{0,\rm ov} = 0.001$, which determines how deep within the convective zone, in terms of pressure scale height, the mixing coefficient is considered before extending beyond the zone.
To mix regions that exhibit an inversion in the mean molecular weight, such as those formed during mass accretion, we use the thermohaline mixing formalism by \cite{Kippenhahn1980}, using an $\alpha_{\rm th} = 666$ coefficient. 

In addition to the aforementioned mixing processes, our models also include rotation, which we assume to be fully synchronized at the start of the evolution, along with the associated rotational mixing mechanisms. We account for angular momentum transport and chemical mixing induced by rotation, following the formalisms of \cite{Heger2000}. This includes processes such as shear instabilities, meridional circulation and the Goldreich-Schubert-Fricke instability. The inclusion of these effects allows for a more realistic treatment of stellar structure and evolution, particularly in the case of massive stars where rotation plays a crucial role in shaping the internal dynamics and surface properties \citep{Maeder2008,Renzo2021,Tauris2023}. After \cite{Heger2000}, we set the diffusion coefficient for rotational mixing to $D_{\rm rot} = 1/30$. 

We also account for the effects of stellar winds, which play a significant role in the evolution of massive stars. Our models incorporate wind mass loss rates based on the prescriptions of \cite{Vink2001}, which depend on stellar parameters such as luminosity and metallicity. While Vink’s formalism is applicable at high temperatures ($T_{\rm eff} > 10\,000,K$), for lower temperatures we supplement the wind models with the prescriptions of \cite{Reimers1975} for the red giant branch (RGB) phase and \cite{Bloecker1995} for the asymptotic giant branch (AGB) phase. For these prescriptions, we adopt efficiency coefficients of $\eta_{\rm V} = 0.1$, $\eta_{\rm R} = 0.5$, and $\eta_{\rm B} = 0.1$, respectively.

For large-scale effects, we used the \cite{Kolb1990} mass transfer prescription and assumed a constant eccentricity throughout the system’s evolution, i.e. $e = 0$. Additionally, we included angular momentum losses via mass loss, gravitational waves, and tidal L-S coupling. Angular momentum accretion was treated according to the formalism of \cite{deMink2013}. We did not, however, account for magnetic braking.

\begin{table}
\caption{A summary of considered parameter ranges during the evolutionary modelling of the V1216\,Sco system.}
\label{tab:MESA_parameter_ranges}
\centering 
\setlength{\tabcolsep}{0pt} 
\begin{tabular}{lccc} 
\hline\hline 
    Parameter   & Range   & Accuracy     \T\B\\ 
    \hline 
    Initial orbital period, $P_{\rm ini}$ \hspace{2pt}[d]       & 1 - 7   & $10^{-5}$  \T\\
    Initial masses, $M_{\rm don/acc,ini}\,\rm[M_{\odot}]$       & 3 - 15  & $10^{-3}$  \\
    Metallicity, $Z$                                            & 0.001 - 0.030 & $10^{-3}$  \\
    Convective-core overshooting, $f_{\rm ov}$                  & 0.001 - 0.040  & $10^{-3}$  \\
    Mixing-length theory parameter, $\alpha_{\rm MLT}$          & 0.5 - 1.5   & $10^{-1}$  \\
    Fraction of mass lost during MT, $\beta$                    & 0.0  - 0.3   & $10^{-1}$  \B\\
    \hline
\hline 
\end{tabular}
\tablefoot{The following columns contain: a description of the parameters, their minimum and maximum values, and the accuracy with which the parameters were drawn from the uniform distributions.}
\end{table}

Including rotation in binary modelling presents a significant challenge, both from a theoretical and numerical perspective. Even the slightest accretion of material that carries angular momentum from the donor can efficiently spin up the accretor, often halting accretion at around 10\% of the initial mass \citep[e.g.,][]{Packet1981}. Rotationally boosted mass loss and magnetic braking are amongst mechanisms that can prevent the star from reaching near-critical rotational velocities, which would otherwise cause the star to break apart \citep[e.g.,][]{deMink2013,Langer2012,UdDoula2009}. However, while magnetic braking is commonly discussed, its implementation and effects in massive stars, where radiative zones dominate, remain uncertain \citep[e.g.,][]{Petit2013,Maeder1999,Maeder2000,Meynet2003}. These uncertainties raise concerns about how well mass accretion in rotating systems is understood.
As a result of these ambiguities, our modelling was not intended to produce highly accurate stellar evolution tracks, but rather to explore the potential formation pathways of systems similar to V1216 Sco. 

We approached the evolutionary modelling in a systematic and iterative manner. Initially, we established a comprehensive grid of 5\,000 models, calculated based on the parameter ranges outlined in Table \ref{tab:MESA_parameter_ranges}. This extensive grid served as a foundation for our analysis, allowing us to explore a wide spectrum of stellar configurations.
From this initial grid, we carefully selected models whose final masses and radii correspond to the central value of the measured orbital period, as detailed in Table \ref{tab:WD_photometric_parameters}. We allowed the model values to deviate by up to 50\% from these observed values, in order to extract a sample of models that enabled us to constrain the ranges of the initial parameters. This approach led to the creation of additional grids, incorporating models with more stringent error thresholds of 30\% and 15\%, respectively. Each of these new grids also consisted of 5\,000 models, allowing us to incrementally hone in on the best-fitting stellar parameters while maintaining a robust sampling of the parameter space.

We identified eight models that fit the observed masses and radii of the components within 15\% errors. The masses of the accretor models range from approximately 10\,\Msun\ to 13\,\Msun, clustering around the observed value of 11.7\,\Msun, while the radii predominantly fall between 7.3\,\Msun\ and 8.3\,\Msun. In contrast, the donor masses span from 4\,\Msun\ to 5\,\Msun, aligning with the observed mass of 4.3\,\Msun\ when considering uncertainties and the evolutionary state during mass transfer. 
In some models we were able to match the primary's (models 1,2,4,5) and secondary's (all models) masses, even within 1$\sigma$ errors. The biggest challenge was to fit the radii of the components, as only one of the models (model 8) matches the primary with 1$\sigma$ and one (model 6) with 3$\sigma$. The rest of the models indicate radii too small to fit the observational range of the primary star. Also, none of the models reproduce the radius of the secondary component, even within 3$\sigma$.
The effective temperatures indicate that the accretor maintains temperatures near 22\,500 K, while the donor stabilizes around 18\,000 K, suggesting that the models capture the expected thermal states under mass transfer conditions. 

All of the models point to the phase of rapid spin-up for the mass accreting star, reaching approximately 99\% of the critical rotational velocity, a common trait in binary systems undergoing substantial mass exchange. This outcome aligns with the measured radius of the donor star (see Table \ref{tab:WD_absolute_parameters}), as donor stars are typically expected to exhibit an inflated radius during the mass transfer phase, which should subsequently decrease once the transfer ceases as they detach from their critical Roche lobes and regain thermal equilibrium \citep[e.g.,][]{Paczynski1971,Soberman1997,Deschamps2013}.
However, this result contradicts observations, which show significantly lower rotational velocities, around 20\% of the critical value. The secondary star, according to the models, rotates at moderate velocities between 10 - 100\, km\,s$^{-1}$, what is in good agreement with the observed values (see Table \ref{tab:spectroscopy_parameters}).
The mass accretion, that causes the star to gain rapid rotation causes also the growth of the convective core mass. During the evolution, the mass accreting star reaches $1.5 - 3.0$\,\Msun\ convective core, corresponding to approximately 10\%-20\% growth compared to the pre-mass transfer stage.
Despite some scatter among models and inherent uncertainties, the models consistently suggest an estimated system age of approximately 15 – 30\,Myr.

\begin{figure*}[htbp]
    \centering
    \includegraphics[width=0.45\textwidth]{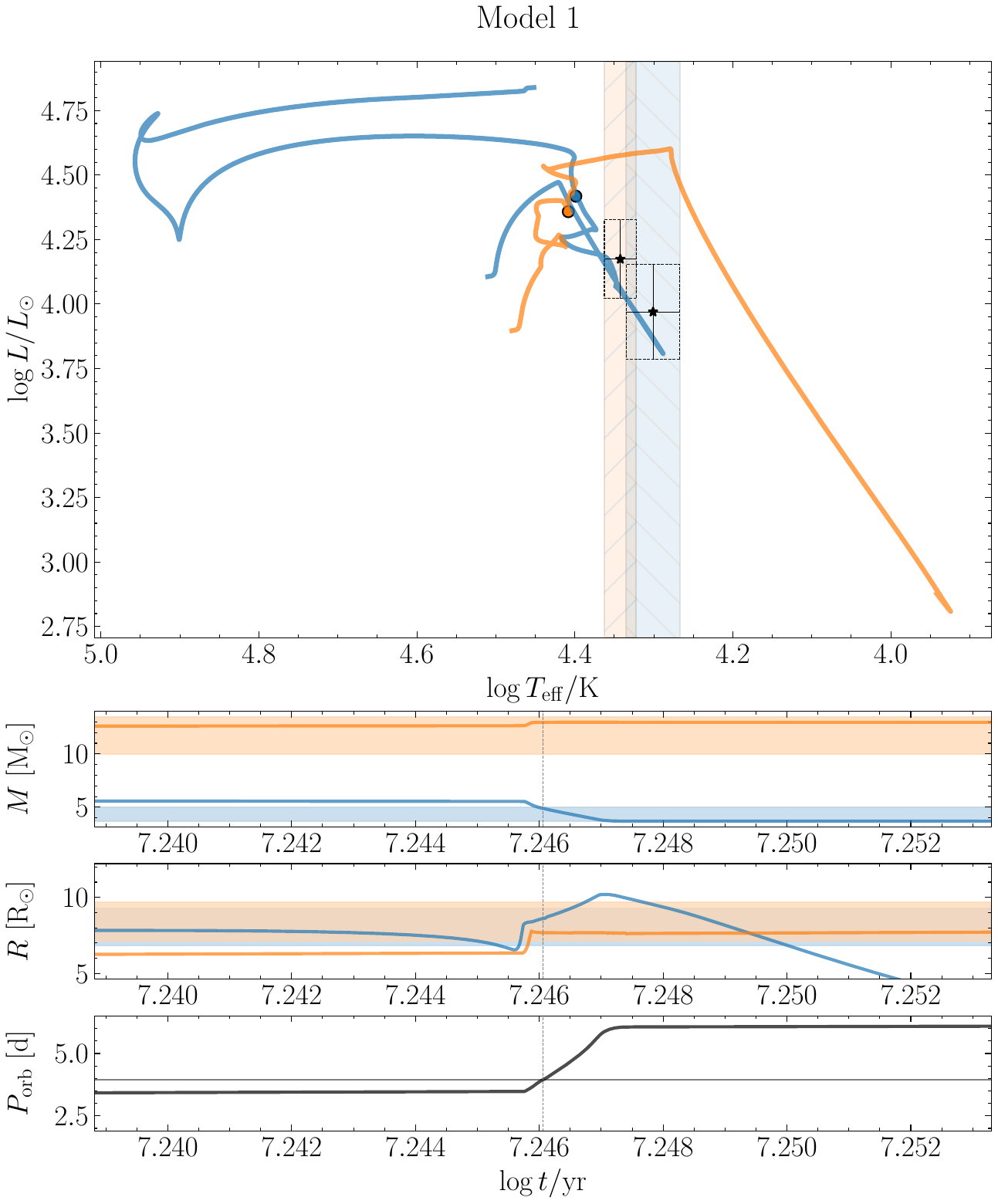} 
    \includegraphics[width=0.45\textwidth]{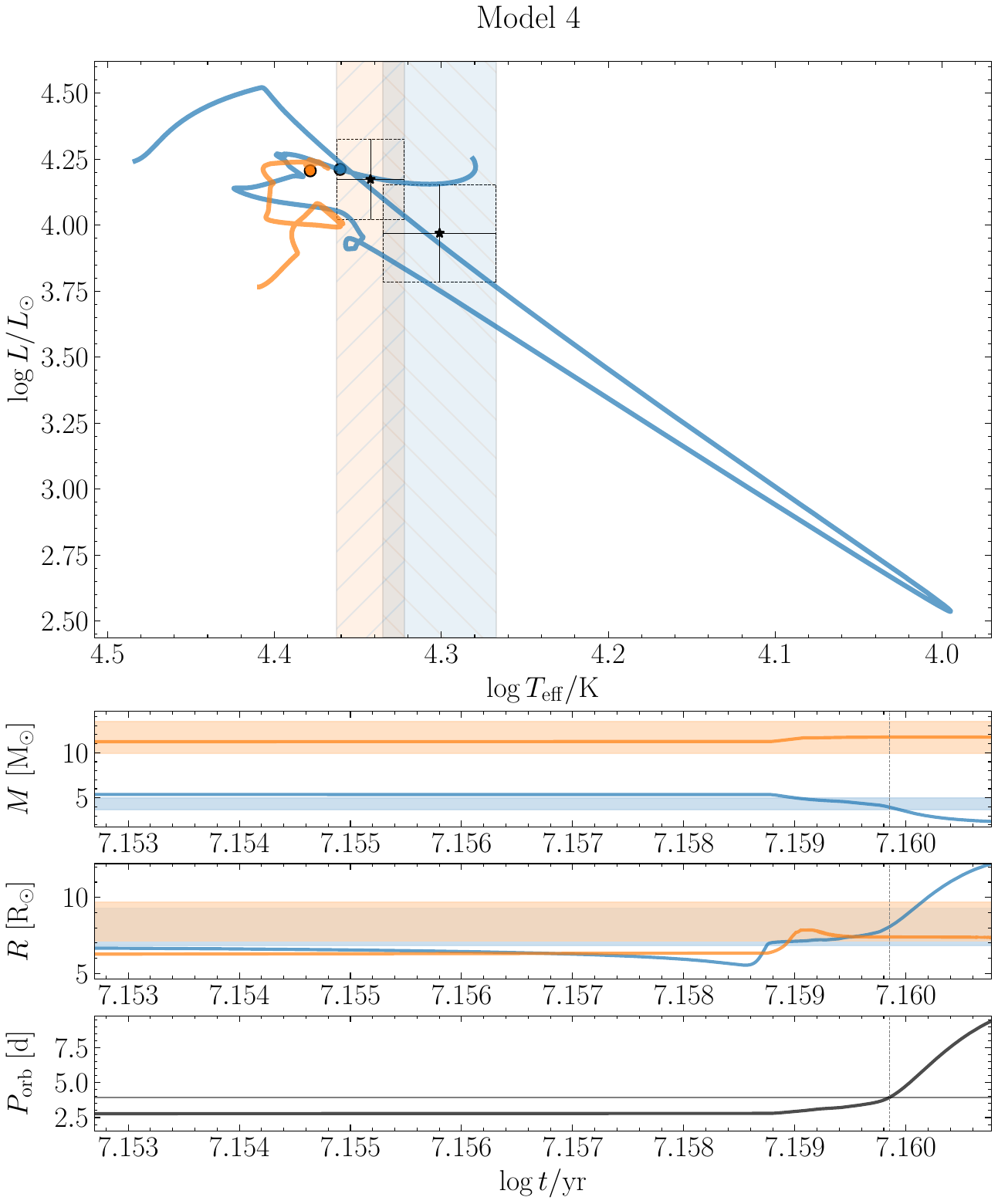}  \\
    \vspace{15pt}
    \includegraphics[width=0.45\textwidth]{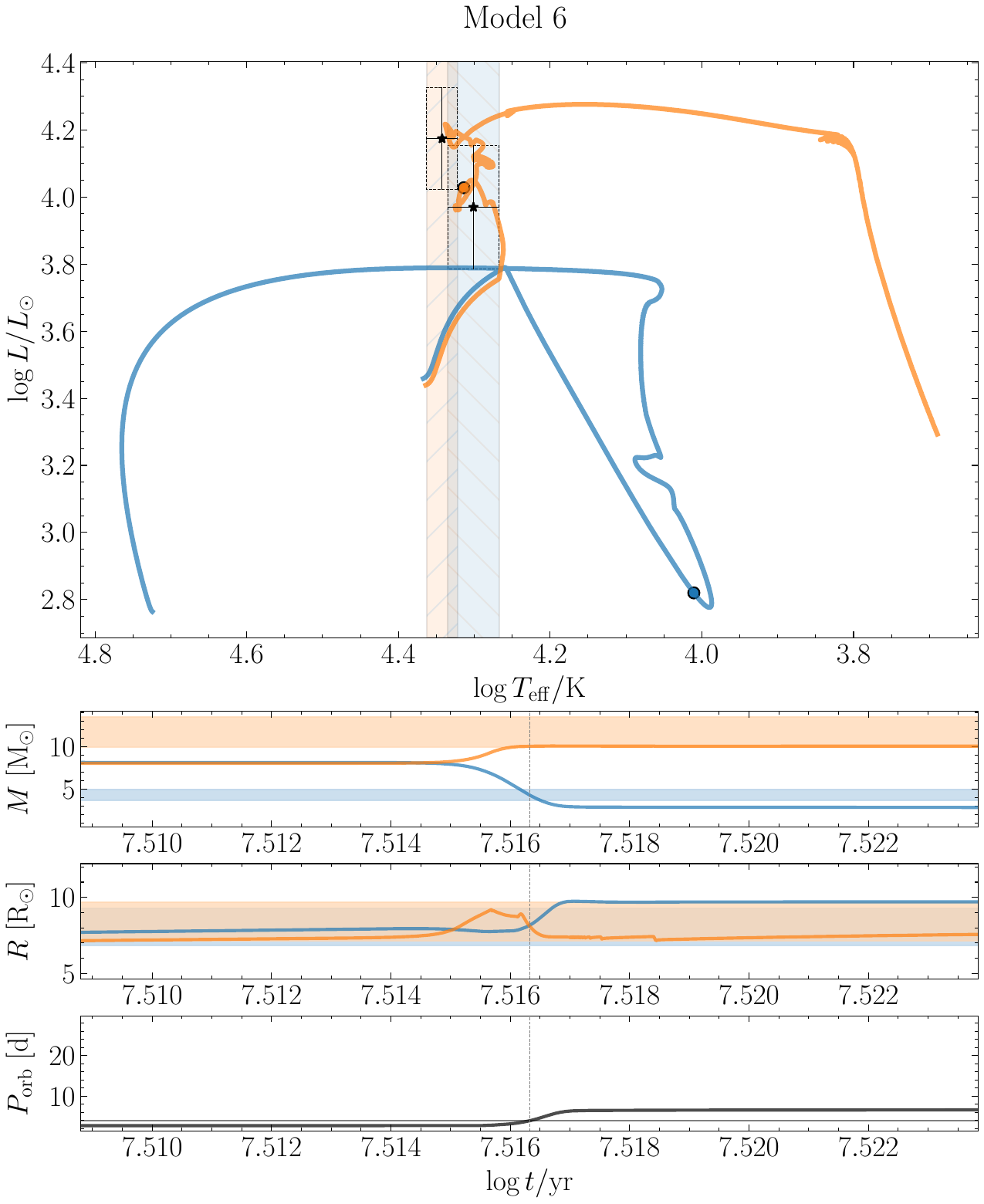} 
    \includegraphics[width=0.45\textwidth]{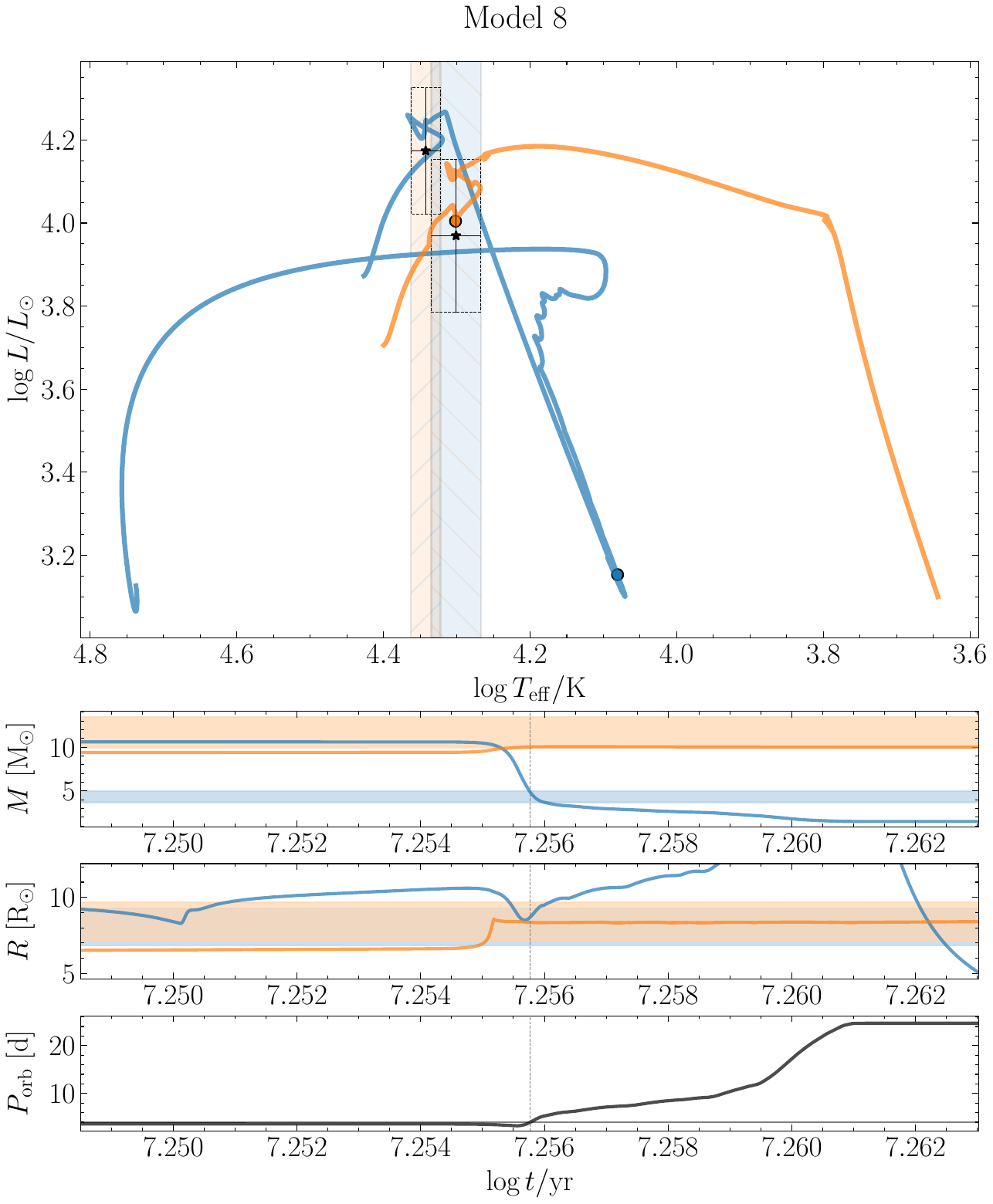} 
    
    \caption{The HR diagrams and the evolution of masses, radii, and orbital period for models 1, 4, 6 and 8 (see Table \ref{tab:MESA_fitting_models}).}
    \label{fig:MESA}
\end{figure*}

The selected models (number 1, 4, 6 and 8, as in Table \ref{tab:MESA_fitting_models}) are shown in Figure \ref{fig:MESA}. Each model is presented with the respective set of evolutionary tracks (blue for donor and orange for acceptor). The following top to bottom panels show the underlying evolution on the HR diagram along with the evolution of masses, radii and the orbital period. The HR diagram has the observed positions of components marked with the 1$\sigma$ error boxes with the vertical, hatched regions showing the spectroscopically determined ranges of effective temperatures for each component.
The horizontal regions on the following panels mark the observed ranges of respective parameters with the 15\% errors while the vertical grey lines indicate the position of model with the observed value of orbital period. This model is marked on each evolutionary track in the HR diagram with orange and blue dots for the acceptor and donor star, respectively.

The reason why we failed in finding a model that fits the observed masses and radii within the measured errors, as we believe, may lie in including rotation in our modelling. Rotation must be included to provide a more realistic theoretical treatment of evolution, but it also introduces an additional level of degeneracy and ambiguity to the solution, making it harder to reconcile all the parameters.
However, our models may prove helpful in understanding the evolutionary channels that lead to creating V1216\,Sco alike systems. Our calculations suggest that \system\ was formed most probably in a case A mass transfer scenario \citep{Kippenhahn1967}, i.e. before the donor left the main sequence, with initial orbital periods between 2.3\,d and 2.9\,d. Only one model (model 8) shows significantly larger value of the initial orbital period, i.e. 3.7\,d. This model reaches the observed state of the system with case B Roche lobe overflow. The initial masses of the components may have been very similar to each other, i.e. 8--14\,\Msun\ for the donor and 8--11\,\Msun\ for the accretor, resulting in an initial mass ratio close to unity. Our models do not favour any particular values of $\alpha_{\rm MLT}$, overshooting, nor the metallicities. However, for galactic B-type stars the latter should not differ much from the solar value, according to \cite{Nieva2012}.

\begin{table*}
\caption{The results of the evolutionary computations summarising the initial and final parameters of the models that best reproduce the system in its present state.}
\label{tab:MESA_fitting_models}
\centering 
\setlength{\tabcolsep}{4pt} 
\begin{tabular}{cccccccc|ccccccc} 
\hline\hline 
    &\multicolumn{7}{c}{\textsc{Initial parameters}} & \multicolumn{7}{c}{\textsc{Final parameters}} \T\\
     Mod.\,\# &$P_{\rm ini}$ & $M_{\rm don,ini}$ & $M_{\rm acc,ini}$ & $Z$  & $\alpha_{\rm MLT}$ & $f_{\rm ov}$ & $\beta$ & $M_{\rm acc}$ & $R_{\rm acc}$ & $\log T_{\rm eff}^{\rm acc}$ & $M_{\rm don}$ & $R_{\rm don}$ & $\log T_{\rm eff}^{\rm don}$ & $\log \rm Age$ \B\\
    \hline 
    1 & 2.33308 & 12.946 & 10.969 & 0.002 & 1.5 & 0.024 & 0.10 & 12.960 & 7.676 & 4.408 & 4.914 & 8.585 & 4.397 & 7.25 \\
    2 & 2.50629 & 8.759 & 8.555 & 0.023 & 0.2 & 0.004 & 0.12 & 10.613 & 7.518 & 4.331 & 4.619 & 8.277 & 4.059 & 7.37 \\
    3 & 2.57735 & 8.812 & 8.175 & 0.017 & 0.3 & 0.036 & 0.16 & 10.292 & 7.343 & 4.324 & 4.714 & 8.170 & 4.162 & 7.45 \\
    4 & 2.61577 & 13.997 & 9.737 & 0.029 & 1.4 & 0.017 & 0.10 & 11.736 & 7.390 & 4.379 & 3.974 & 8.038 & 4.362 & 7.16 \\
    5 & 2.78764 & 9.412 & 9.050 & 0.026 & 0.1 & 0.026 & 0.01 & 11.393 & 8.037 & 4.349 & 4.996 & 8.510 & 4.149 & 7.34 \\
    6 & 2.82684 & 8.151 & 8.038 & 0.013 & 0.3 & 0.018 & 0.01 & 10.026 & 8.143 & 4.313 & 4.352 & 8.141 & 4.012 & 7.52 \\
    7 & 2.88185 & 10.421 & 9.216 & 0.027 & 0.4 & 0.022 & 0.10 & 10.406 & 7.685 & 4.325 & 4.602 & 8.301 & 4.178 & 7.25 \\
    8 & 3.70749 & 10.634 & 9.404 & 0.026 & 0.2 & 0.009 & 0.02 & 10.041 & 8.345 & 4.302 & 4.963 & 8.654 & 4.079 & 7.26 \\
    \hline
\hline 
\end{tabular}
\tablefoot{$^{\dagger}$ From $\log L/L_{\odot} = 4 \log(T_{\rm eff}/T_{\rm eff \odot}) + 2 \log(R/R_{\odot})$}
\end{table*}

\section{Pulsational modelling}
\label{sec:puls_modelling}

As we were unable to identify models that simultaneously reproduce the observed masses and radii within their measured uncertainties, we limited our seismic analysis to investigating mode instabilities within the observed frequency range. This approach aims to explore whether the pulsational properties of the models provide additional constraints on the system’s evolutionary state.

For the models found in the previous section (see Section \ref{sec:modelling:evolution_modelling} and Table \ref{tab:MESA_fitting_models}) we computed non-adiabatic pulsations for $\ell = 0,1,2$ modes using \textsc{gyre} \citep[version 7.1,][]{GYRE-Townsend2013,GYRE-Townsend2018,GYRE-Goldstein2020}. Since our equilibrium models account for the effects of differential rotation, we included rotational splitting by considering azimuthal orders in the range $m \in [-\ell, \ell]$.

Our prime goal was to conduct a pulsational stability analysis for the frequencies found in the V1216\,Sco light curve (see Table \ref{tab:frequencies}), noting that none of these frequencies have assigned $\ell$ and $m$ values. The stability parameter $\eta$, introduced by \cite{Stellingwerf1978}, is a normalized work integral computed over the pulsational cycle. If $\eta > 0$ for a given mode, the driving mechanism overcomes damping, and the mode is excited; otherwise, the mode is suppressed in the stellar interior.

\begin{figure}[htbp]
    \centering
    \includegraphics[width=0.48\textwidth]{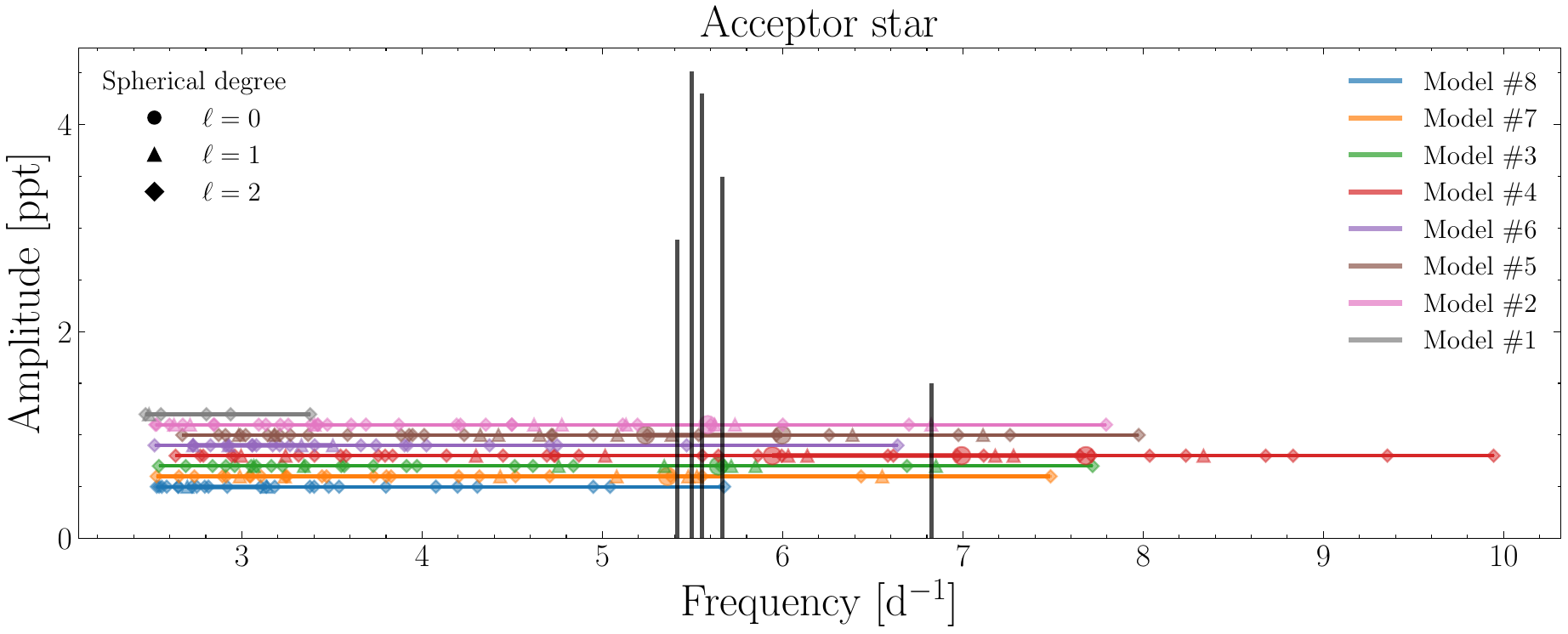} 
    \caption{Rotationally split unstable $\ell=0,1,2$ modes calculated for accretor evolutionary models summarised in Table \ref{tab:MESA_fitting_models}.}
    \label{fig:MESA_puls}
\end{figure}

We compared the ranges of excited modes for both binary components with the observed frequencies for each model we have obtained in the previous section. We found that only three models for the accretor fail to cover the observed frequency range with unstable modes, while the remaining models successfully reproduce this range with both low-order p-modes and low-order g-modes. These are models no. 1, 6 and 8 (see Table \ref{tab:MESA_fitting_models}). None of the models for the donor show any instabilities, except for model no. 4, which exhibits unstable high-order g-modes between 1\,\cpd and 2\,\cpd. This may indirectly suggest that the observed frequencies originate from the more massive component. This conclusion could not be confidently drawn based solely on the components’ flux ratio, as the primary component contributes approximately $\sim 55\%$ of the total flux. However, the observed masses suggest that the donor component would pulsate as an SPB type rather than as a $\beta$\,Cep pulsator. The results are shown in Figure\,\ref{fig:MESA_puls}, where we plot the independent pulsational frequencies from \TESS sector 66 alongside the ranges of rotationally split unstable modes for each model. 

\section{Conclusions}
\label{sec:conclusions}

In this study, we analysed the binary system V1216\,Sco using \TESS photometry and spectroscopic data from the South African Large Telescope High Resolution \'echelle Spectrograph. We aimed to refine the system’s orbital parameters, component properties, and pulsational characteristics through detailed modelling of both photometric and spectroscopic data. We also attempted to find evolutionary models that fit the observations.

The \TESS\ photometry, consisting of around 12\,500 data points from three sectors, was used to construct the system’s light curve, which exhibited strong pulsations that were accounted for during the orbital modelling process. Spectroscopic data, consisting of nine spectra obtained in 2022, allowed us to derive the radial velocities of the system and perform a detailed spectral analysis.

The RV data were used to determine the orbital parameters, and the disentangled spectra provided the atmospheric properties of the two components of the binary. Using the Wilson-Devinney code, we performed a comprehensive orbital modelling of the system, fitting both the radial velocities and the light curve data. The modelling results confirmed that the system is very close to circular, with both stars having similar radii despite a significant mass difference. The primary star has a mass of 11.72\,\Msun\ and the secondary star has a mass of 4.35\,\Msun. The secondary’s large radius, close to its Roche lobe, supports the conclusion that the system is undergoing or has recently undergone mass transfer.

A Fourier analysis of the residual light curve revealed pulsations in the system, with frequencies covering the typical $\beta$\,Cephei range. We found 8 frequencies in the range $\sim 5.4$\,\cpd\ to  $\sim 11.1$\,\cpd, from which 5 strongest seem to be independent, within the criteria we adopted.


In light of the large radius of the secondary component, which could not be easily explained by a single-star evolution model, we hypothesized that V1216\,Sco is a product of a system that has undergone mass transfer. To further investigate this, we employed the \textsc{mesa-binary} code to model the binary system’s evolution. We constructed an extensive grid of evolutionary models with varying initial conditions, such as the masses, orbital period, and metallicity of the components, as well as the fraction of mass lost during the mass transfer phase. Our results indicate that V1216\,Sco was likely formed in a case A mass transfer scenario, where mass transfer occurred before the donor left the main sequence, with initial orbital periods between 2.3\,d and 2.9\,d and initial masses ranging from 8\,\Msun\ to 14\,\Msun\ for the donor and 8\,\Msun\ to 11\,\Msun\ for the accretor, resulting in a near-unity mass ratio. These models suggest that the system underwent significant mass transfer in the past, likely inflating the radius of the former donor, now the secondary component, to its current state.

Furthermore, our evolutionary models suggest that the mass accreting star in V1216\,Sco should have reached velocity near 99\% of the critical rotational speed, due to substantial mass exchange and angular momentum accretion. Despite this, observational data show much lower rotational velocity for the accretor, suggesting that rotational effects may be suppressed or that our models may have overestimated the rotational velocities due to the complexities of mass transfer dynamics and angular momentum conservation. The mass accretion also contributed to the growth of the convective core of the accretor, which reached up to 3.0\,\Msun, during the mass transfer phase. The models predict an age of approximately 15 – 30\,Myr for the system, although some scatter among the models and uncertainties in key parameters make this estimate less precise. Using evolutionary and pulsational models we were able to reproduce the observed frequency range with unstable $\ell=0,1,2$ modes.
In the future, we plan to collect multicolour photometry for V1216 Sco, which will allow us to perform mode identification. This will help us further constrain the evolutionary models, providing asteroseismic modelling and more detailed understanding of the system’s parameters.

Overall, while the evolutionary models did not precisely match the observed masses and radii within the error margins, they provided valuable insights into the possible formation and evolutionary pathways of systems like V1216\,Sco. The results suggest that the mass transfer played a significant role in shaping the observed properties of both stars. Further refinements in the modelling of rotation and mass accretion processes could help better constrain the evolutionary history of such systems.

\begin{acknowledgements}
\noindent This work was supported by the Polish National Science Centre (NCN), grant number 2021/43/B/ST9/02972.

AM thanks John Southworth, Andrzej Pigulski, Slavek Ruciński and Andrej Pr\v{s}a for inspiring discussions regarding the nature of V1216 Sco light curve, both from the observational and the modelling perspective.

This paper includes data collected by the \textit{TESS} mission. Funding for the \textit{TESS} mission is provided by the NASA Explorer Program. Funding for the \textit{TESS} Asteroseismic Science Operations Centre is provided by the Danish National Research Foundation (Grant agreement no.: DNRF106), ESA PRODEX (PEA 4000119301) and Stellar Astrophysics Centre (SAC) at Aarhus University. We thank the \textit{TESS} team and staff and TASC/TASOC for their support of the present work. 

Calculations have been carried out using resources provided by Wroc\l aw Centre for
Networking and Supercomputing (http://wcss.pl), grant no. 265. 

All plots in this paper have been created using the \textsc{scienceplots} package by \cite{SciencePlots}.
\end{acknowledgements}

\section*{Data Availability}
The target pixel files were downloaded from the public data archive at MAST. The light curves will be shared upon a reasonable request. 
We make all files needed to recreate our \textsc{mesa-binary} and \textsc{gyre} results publicly available at Zenodo: 
\href{https://www.doi.org/10.5281/zenodo.14198585}{10.5281/zenodo.14198585}.

\section*{Software}
This paper made use of the following codes/packages:
\textsc{lightkurve} \citep{lightkurve2018}, \textsc{spectrum} \citep{Gray1999}, \textsc{iraf/pyraf} \citep{Tody1986, PyRAF2012}, \textsc{rvfit} \citep{Iglesiasetal2015}, \textsc{fd3binary} \citep{Ilijicetal2004}, \textsc{wd} \citep{Wilson1971,Wilson1979,Wilson1990}, \textsc{emcee} \citep{Foreman_Mackey_2013}, \textsc{celerite} \citep{celerite}, 
\textsc{mesa} \citep{Paxton2011,Paxton2013,Paxton2015,Paxton2018,Paxton2019,Jermyn2023}, \textsc{pyMESAreader} (\href{https://billwolf.space/py\_mesa\_reader/index.html} {https://billwolf.space/py\_mesa\_reader/index.html}), \textsc{gyre} \citep{GYRE-Townsend2013,GYRE-Townsend2018,GYRE-Goldstein2020,GYRE-Sun2023}, \textsc{Python NumPy} \citep{NumPy}, \textsc{Python SciPy} \citep{2020SciPy-NMeth}

\begin{appendix}

\section{\textsc{mesa} input physics}
\label{sec:mesa_physics}
The \textsc{mesa} code builds upon the efforts of many researchers who advanced our understanding of physics and relies on a variety of input microphysics data.
The \textsc{mesa} EOS is a blend of the OPAL \citep{Rogers2002}, SCVH \citep{Saumon1995}, FreeEOS \citep{Irwin2004}, HELM \citep{Timmes2000}, and PC \citep{Potekhin2010} EOSes.
Radiative opacities are primarily from the OPAL project \citep{Iglesias1993,Iglesias1996}, with data for lower temperatures from \citet{Ferguson2005} and data for  high temperatures, dominated by Compton-scattering from \citet{Buchler1976}. Electron conduction opacities are from \citet{Cassisi2007}.
Nuclear reaction rates are from JINA REACLIB \citep{Cyburt2010} plus additional tabulated weak reaction rates from \citet{Fuller1985}, \cite{Oda1994} and \cite{Langanke2000}. Screening is included via the prescription of \citet{Chugunov2007}. Thermal neutrino loss rates are from \citet{Itoh1996}.
The \textsc{mesa-binary} module allows for the construction of a binary model and the simultaneous evolution of its components, taking into account several important interactions between them. In particular, this module incorporates angular momentum evolution due to mass transfer.
Roche lobe radii in binary systems are computed using the fit of \citet{Eggleton1983}. Mass transfer rates in Roche lobe overflowing binary systems are determined following the prescriptions of \citet{Ritter1988} and \citet{Kolb1990}.

\section{Corner plots}
\label{sec:corner_plots}
Below, we provide the corner plots that come from the solutions for radial velocity data, photometric modelling, and the derived absolute parameters from the orbital analysis presented in Section\,\ref{sec:modelling:orbital_modelling}.

\begin{figure*}[htbp]
    \centering
    \begin{minipage}{0.48\textwidth}
        \includegraphics[width=\textwidth]{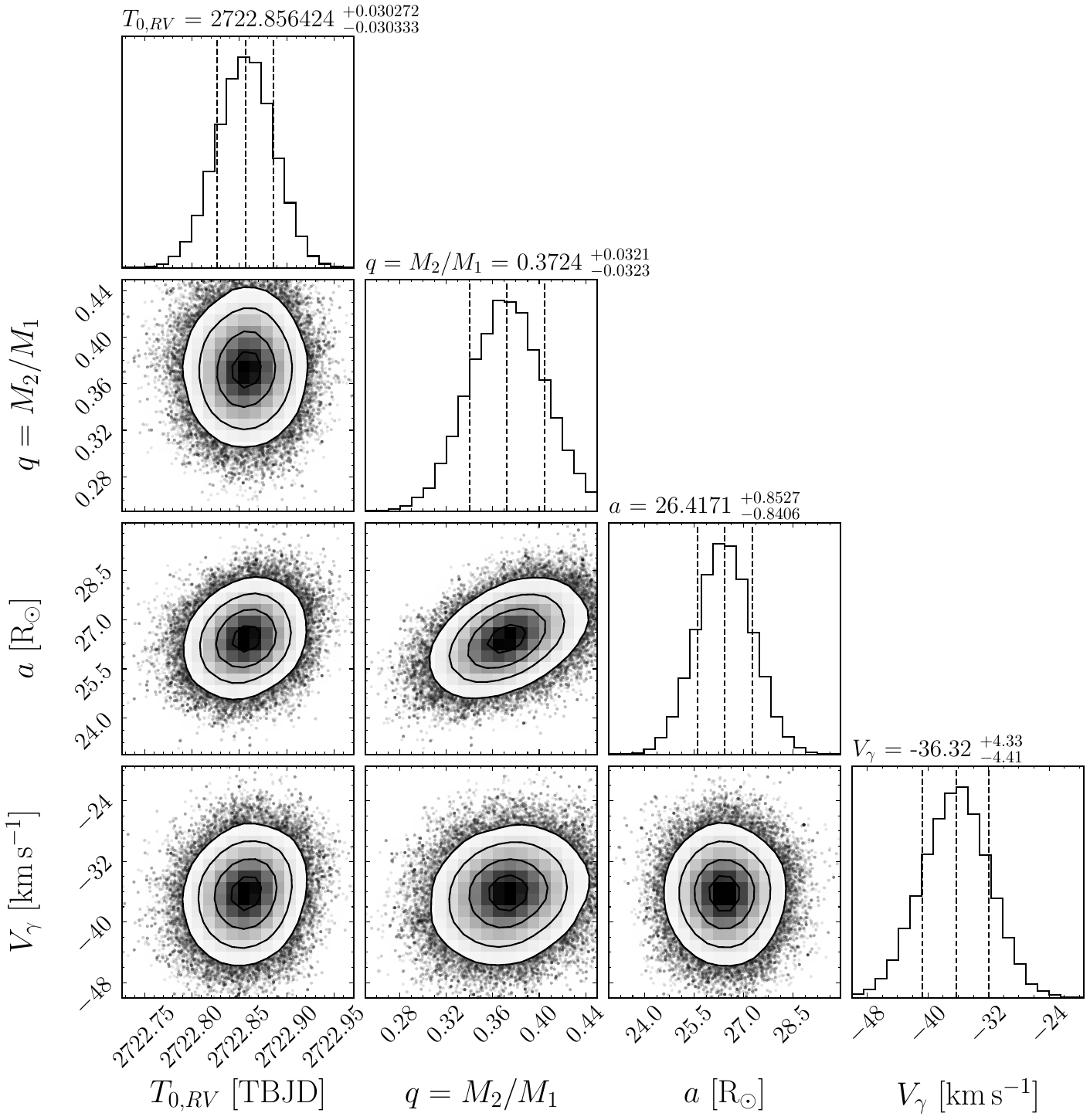} 
        \caption{Corner plots displaying the posterior distributions among different parameters obtained from the radial velocities MCMC sampling. The listed parameter values represent the mode of the posterior distributions, with uncertainties provided as the $16^{\rm th}$ and $84^{\rm th}$ percentiles, corresponding to the $1\sigma$ credible intervals.}
        \label{fig:RV_corner}
    \end{minipage}
    \hfill
    \begin{minipage}{0.48\textwidth}
        \includegraphics[width=\textwidth]{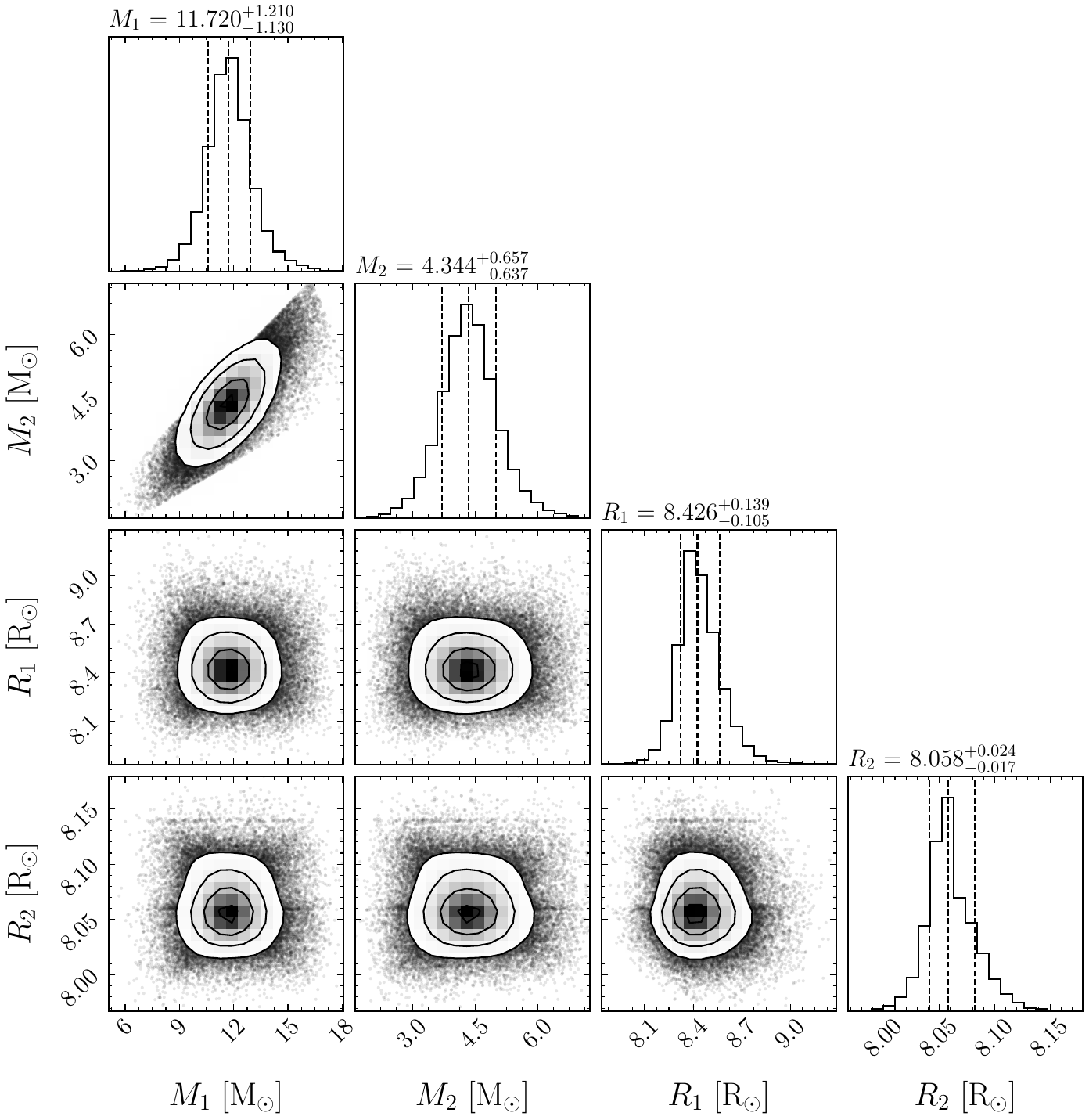} 
        \caption{Corner plots displaying the posterior distributions for masses and radii from the MCMC sampling. The listed parameter values represent the mode of the posterior distributions, with uncertainties provided as the $16^{\rm th}$ and $84^{\rm th}$ percentiles, corresponding to the $1\sigma$ credible intervals.}
        \label{fig:MR_corner}
    \end{minipage}
\end{figure*}

\begin{figure*}[htbp]
    \centering
    \includegraphics[width=\textwidth]{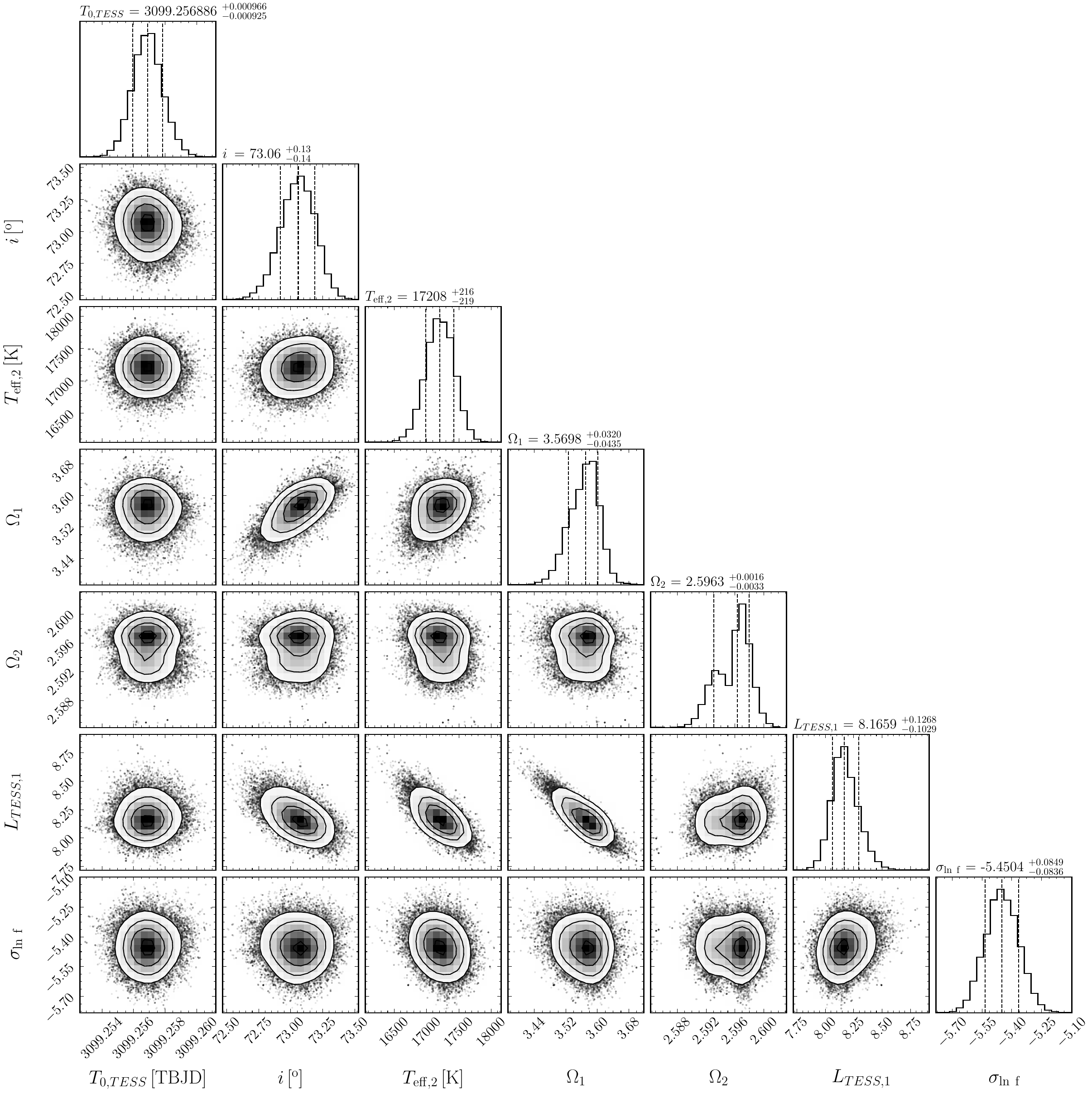} 
    \caption{Corner plots displaying the posterior distributions among different parameters obtained from the light curve MCMC sampling. The listed parameter values represent the mode of the posterior distributions, with uncertainties provided as the $16^{\rm th}$ and $84^{\rm th}$ percentiles, corresponding to the $1\sigma$ credible intervals.}
    \label{fig:LC_corner}
\end{figure*}

\end{appendix}

\bibliography{bibliography}{}
\bibliographystyle{aasjournal}



\end{document}